\documentclass[10pt,preprint]{aastex}

\begin{document}

\shortauthors{Luhman}
\shorttitle{Census of Chamaeleon~I}

\title{A Census of the Chamaeleon~I Star-Forming Region \altaffilmark{1}}

\author{K. L. Luhman}
\affil{Harvard-Smithsonian Center for Astrophysics, 60 Garden St.,
Cambridge, MA 02138}

\email{kluhman@cfa.harvard.edu}

\altaffiltext{1}{Based on observations performed at Las Campanas Observatory.
This publication makes use of data products from the Two Micron All  
Sky Survey, which is a joint project of the University of Massachusetts 
and the Infrared Processing and Analysis Center/California Institute 
of Technology, funded by the National Aeronautics and Space
Administration and the National Science Foundation.}

\begin{abstract}

I present a new census of the members of the Chamaeleon~I star-forming region.
Optical spectroscopy has been obtained for 179 objects 
that have been previously identified as possible members of the cluster, that
lack either accurate spectral types or clear evidence of membership, and 
that are optically visible ($I\lesssim18$).
I have used these spectroscopic data and all other available constraints 
to evaluate the spectral classifications and membership status of a total
sample of 288 candidate members of Chamaeleon~I that have appeared in 
published studies of the cluster.
The latest census of Chamaeleon~I now contains 158 members, 
8 of which are later than M6 and thus are likely to be brown dwarfs.
I find that many of the objects identified as members of Chamaeleon~I in 
recent surveys are actually field stars.
Meanwhile, 7 of 9 candidates discovered by \citet{car02} are confirmed as
members, one of which is the coolest known member of Chamaeleon~I
at a spectral type of M8 ($\sim0.03$~$M_{\odot}$).
I have estimated extinctions, luminosities, and effective temperatures for 
the members and used these data to construct an H-R diagram for the cluster. 
Chamaeleon~I has a median
age of $\sim2$~Myr according to evolutionary models, and hence is similar in
age to IC~348 and is slightly older than Taurus ($\sim1$~Myr).
The measurement of an IMF for Chamaeleon~I from this census 
is not possible because of the disparate methods with which the known 
members were originally selected, and must await an unbiased, magnitude-limited 
survey of the cluster.

\end{abstract}


\section{Introduction}

The targets for observational studies of young stars and brown dwarfs are
selected from surveys for members of nearby star-forming regions. 
The masses for these targets are typically estimated by measuring their spectral
types and interpreting their positions on a Hertzsprung-Russell (H-R) diagram 
with theoretical evolutionary models.
Thus, the success of work on young stars and brown dwarfs relies 
on clear evidence of membership and accurate spectral classifications
for putative members of star-forming populations.

At a distance of 160-170~pc \citep{whi97,wic98,ber99}, 
the Chamaeleon~I cloud complex is one of the nearest major sites of 
active star formation.
A variety of methods have been used to identify members of Chamaeleon~I, 
including monitoring of photometric variability at optical \citep{hof62} 
and infrared (IR) \citep{car02} wavelengths, objective prism spectroscopy at 
H$\alpha$ \citep{hen63,men72,hm73,sch77,har93,com99,com00}, X-ray imaging 
with the {\it Einstein Observatory} \citep{fk89}
and the {\it R\"ontgen Satellite}
\citep{fei93,alc95,alc97,com00}, and near- to mid-IR photometry from 
ground-based telescopes \citep{hjm82,jon85,cam98,ots99,gk01,kg01,per01,car02}, 
the {\it Infrared Astronomical Satellite (IRAS)} 
\citep{bau84,ass90,whi91,pru91,gs92}, and the {\it Infrared
Space Observatory (ISO)} \citep{nor96,per99,per00,com00,leh01}.
To measure spectral types and to check for signatures of youth and membership 
in the resulting candidate members, spectroscopy was employed by those authors
and in subsequent followup work 
\citep{app77,app79,ryd80,app83,wal92,hlf94,law96,cov97,nc99,gp02,gm03,saf03}.

Although Chamaeleon~I has been the target of a large number of studies, 
many of the objects that have been referred to as members of the cluster lack 
either conclusive evidence of membership or accurate spectral classifications. 
In addition, many of the candidates from recent surveys have not been 
observed with spectroscopy.
To address these shortcomings in the current census of Chamaeleon~I, 
I present optical spectroscopy for most of the objects that have been 
previously identified as possible members of the cluster, that lack either
accurate spectral classifications or evidence of membership, and that are 
sufficiently bright ($I\lesssim18$) (\S~\ref{sec:obs}).
I then measure spectral types from these data and use all available 
constraints to evaluate the membership status of most of the 
candidate members of Chamaeleon~I that have appeared in previous work 
(\S~\ref{sec:class}).
The implications of this new census for recent searches for members of the 
cluster are described (\S~\ref{sec:imp}). Finally, I estimate extinctions, 
luminosities, and effective temperatures for the known members and
use these data to construct an H-R diagram for Chamaeleon~I (\S~\ref{sec:prop}).

\section{Observations}
\label{sec:obs}

\subsection{Selection of Targets}
\label{sec:targets}

Most of the sources toward Chamaeleon~I that are discussed here 
have multiple identifications from published surveys. 
In this work, I use the names adopted by \citet{car02}. 
A comprehensive list of cross identifications was provided in that study.

Through near-IR photometric monitoring of a $0\fdg72\times6\arcdeg$
area toward Chamaeleon~I, \citet{car02} recently discovered 10 objects 
that exhibited variability or excess emission that were suggestive of youth and 
membership in the star-forming region. I obtained spectra for nine of these 
sources. The remaining object, CHSM~17388, was not detected
by the optical guide camera and therefore was not observed. In addition to 
these candidates, \citet{car02} compiled astrometry, near-IR photometry, 
and membership information for most of the potential members reported in 
previous studies. I selected for spectroscopy 147 of these objects, which
included all of the optically visible sources ($I\lesssim18$) that lacked
either accurate spectral types or conclusive evidence of membership. 
In their compilation, \citet{car02} excluded most of the candidates reported
by \citet{cam98} and \citet{gk01}. 
I selected eight candidates from \citet{gk01} and the 13 optically
visible candidates that were observed spectroscopically by \citet{gm03},
which were originally identified by \citet{cam98}. I also observed Cam2-29
and YY~Cha.

\subsection{Spectroscopy}

Table~\ref{tab:log} summarizes long-slit optical spectroscopy 
of the 179 targets in Chamaeleon~I from the previous section 
and a sample of standard stars for spectral classification.
These observations were performed with the Boller and Chivens Spectrograph 
(B\&C) and the Low Dispersion Survey Spectrograph (LDSS-2) on 
the Magellan~I and II 6.5~m telescopes at Las Campanas Observatory.
The B\&C spectra with the 600 and 1200~l~mm$^{-1}$ 
gratings covered 6000-9200 and 6200-7800~\AA, respectively,
while the data from LDSS-2 covered 4800-9200~\AA.
All data were obtained with the slit rotated to the parallactic angle. 
The exposure times ranged from 10 to 1200~s. 
After bias subtraction and flat-fielding,
the spectra were extracted and calibrated in wavelength with arc lamp data.
The data were corrected for the sensitivity functions of the detectors,
which were measured from observations of spectrophotometric standard stars.
When available, $i$ measurements from the DENIS Second Release were used to
flux calibrate the spectra that extended to 9200~\AA. These calibrated spectra
were then taken as flux standards for the 13 sources that lacked $i$ data.
The higher-resolution spectra at 6200-7800~\AA\ were calibrated 
by scaling them to the low-resolution data.

\begin{figure} 
\plotone{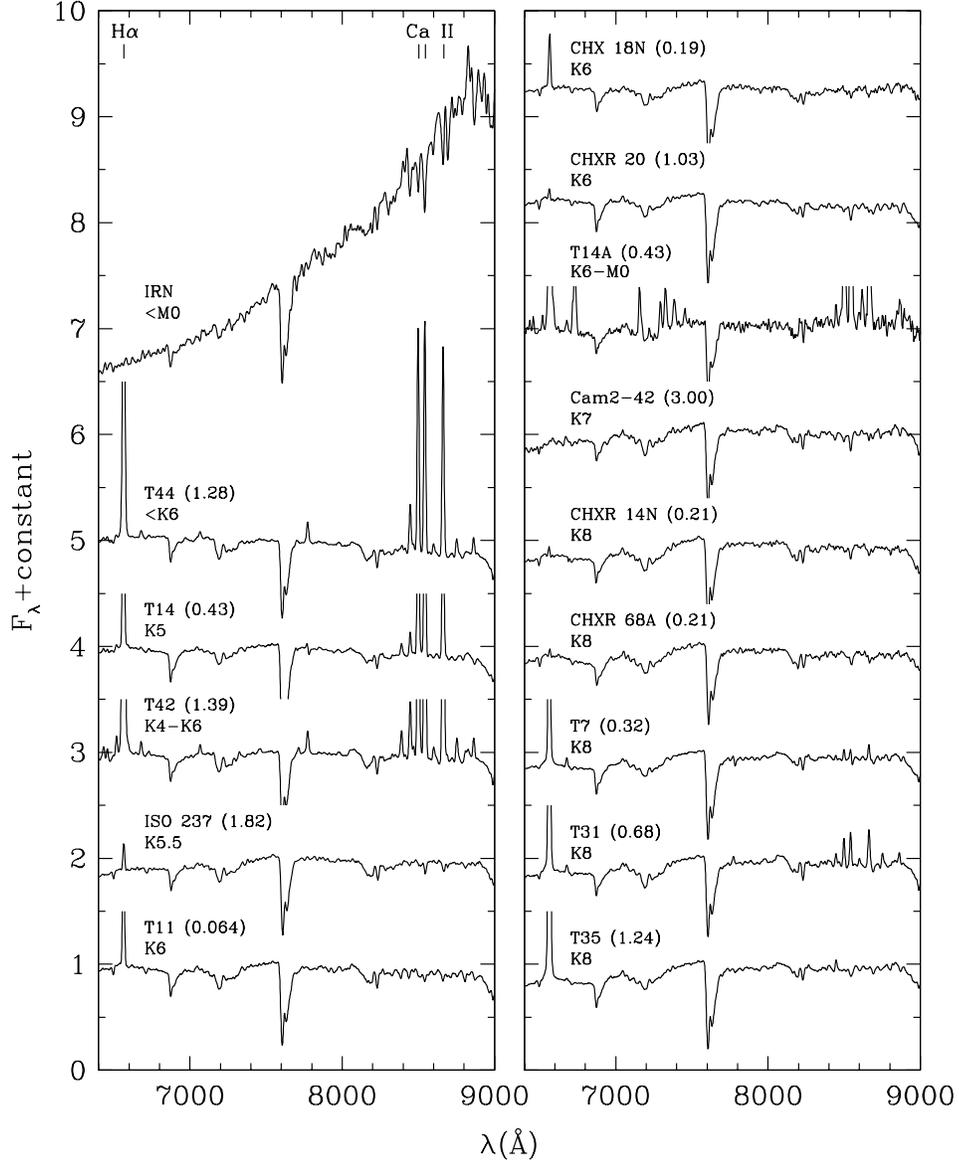}
\caption{
Low-resolution spectra of objects that are classified as members of the 
Chamaeleon~I star-forming region (\S~\ref{sec:confirm}).
The spectra have been corrected for extinction (except for IRN), which 
is quantified in parentheses by the magnitude difference of the reddening 
between 0.6 and 0.9~\micron\ ($E(0.6-0.9)$, \S~\ref{sec:ext}).
The data are displayed at a resolution of 13~\AA\ and are normalized at 
7500~\AA.}
\label{fig:m1}
\end{figure}

\begin{figure}
\plotone{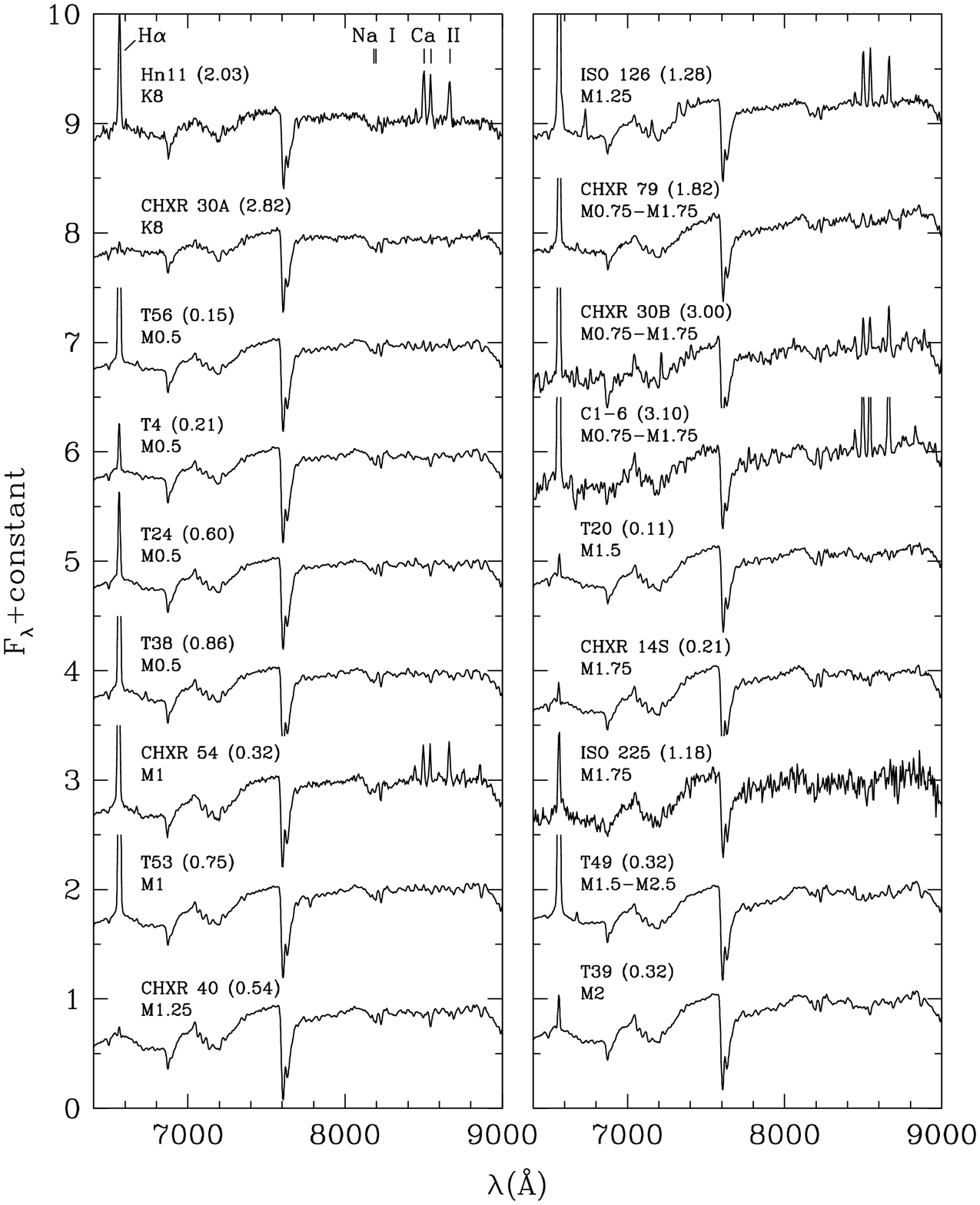}
\caption{Same as Figure~\ref{fig:m1}.}
\label{fig:m2}
\end{figure}

\begin{figure}
\plotone{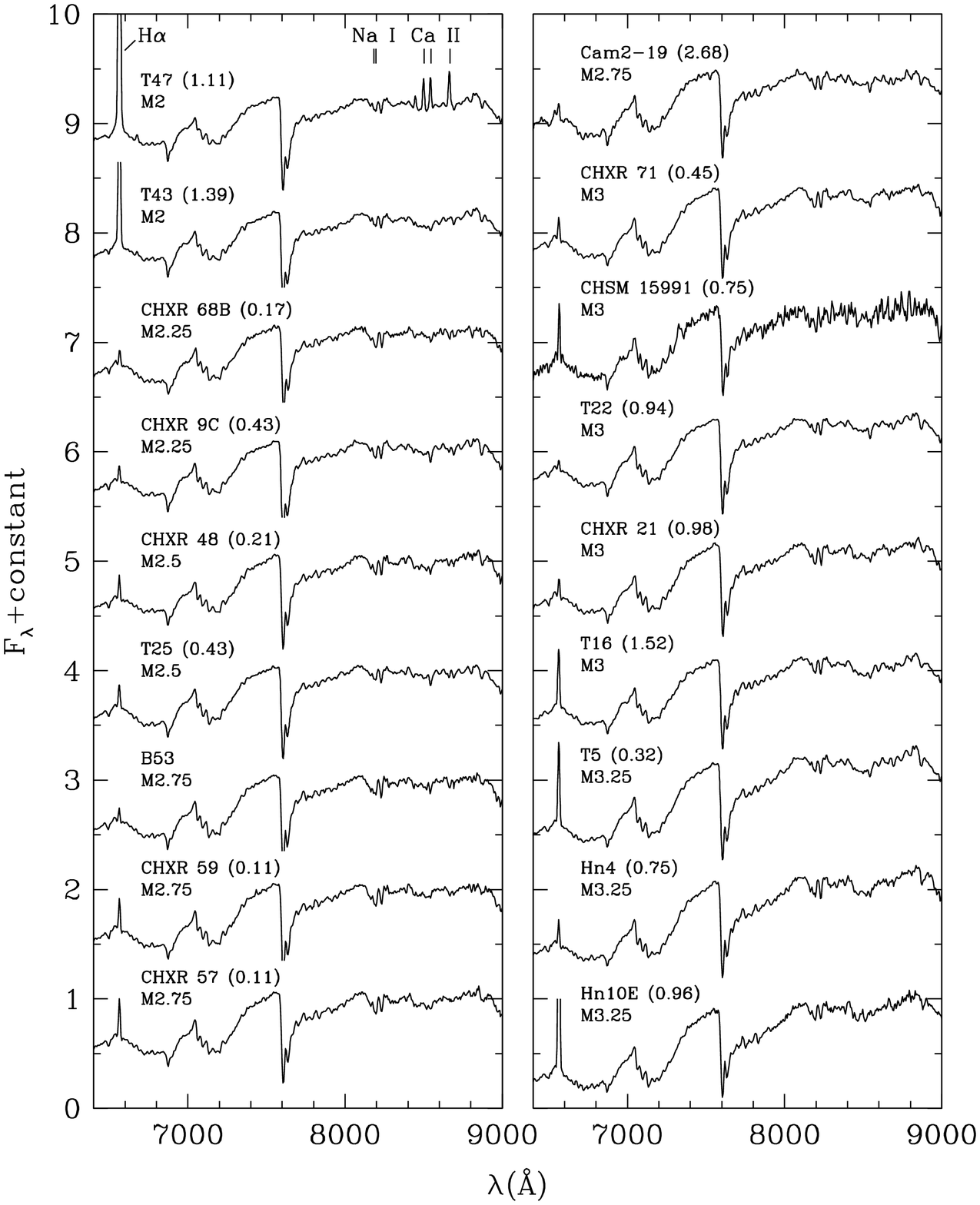}
\caption{Same as Figure~\ref{fig:m1}.}
\label{fig:m3}
\end{figure}

\begin{figure}
\plotone{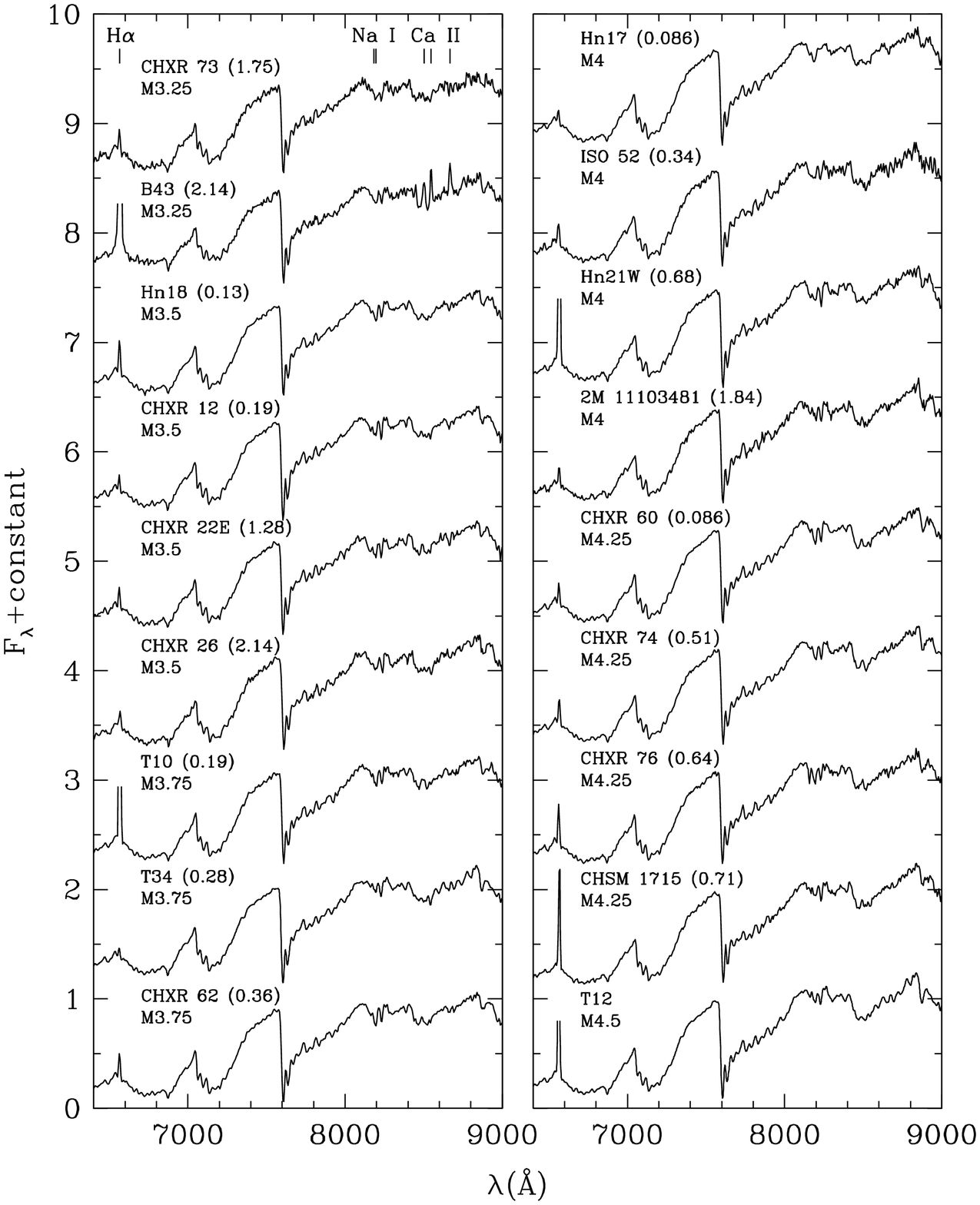}
\caption{Same as Figure~\ref{fig:m1}.}
\label{fig:m4}
\end{figure}

\begin{figure}
\plotone{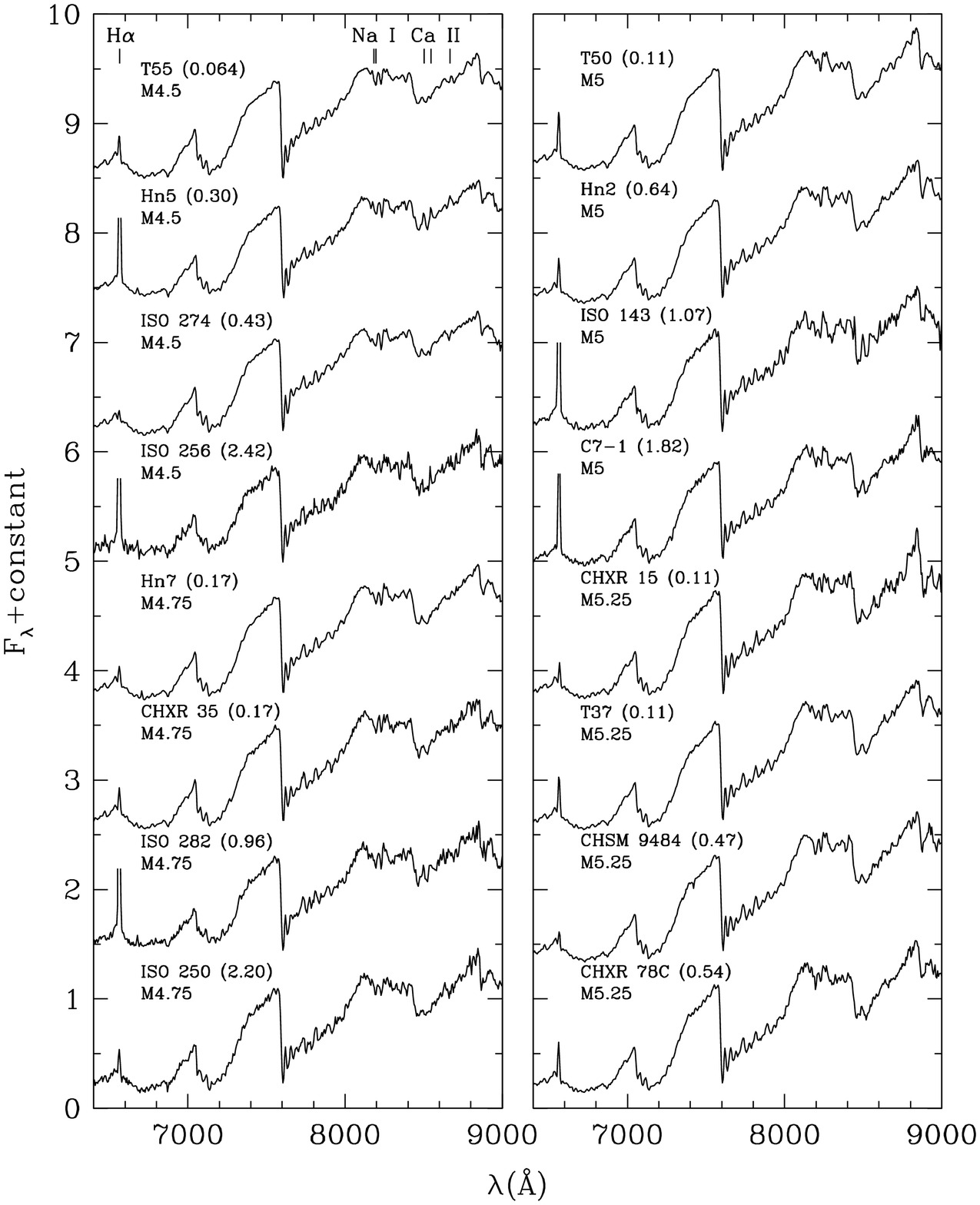}
\caption{Same as Figure~\ref{fig:m1}.}
\label{fig:m5}
\end{figure}

\begin{figure}
\plotone{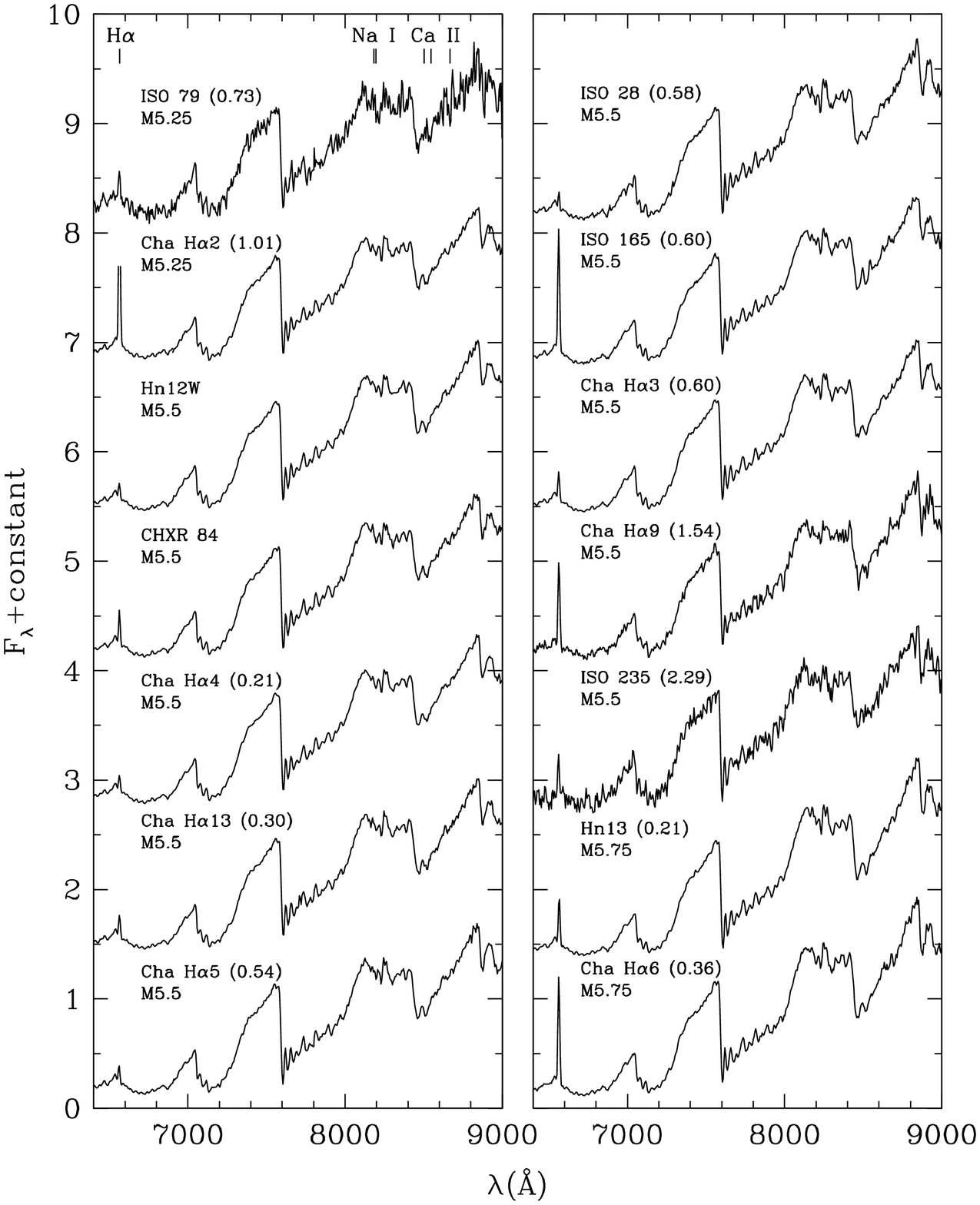}
\caption{Same as Figure~\ref{fig:m1}.}
\label{fig:m6}
\end{figure}

\begin{figure}
\plotone{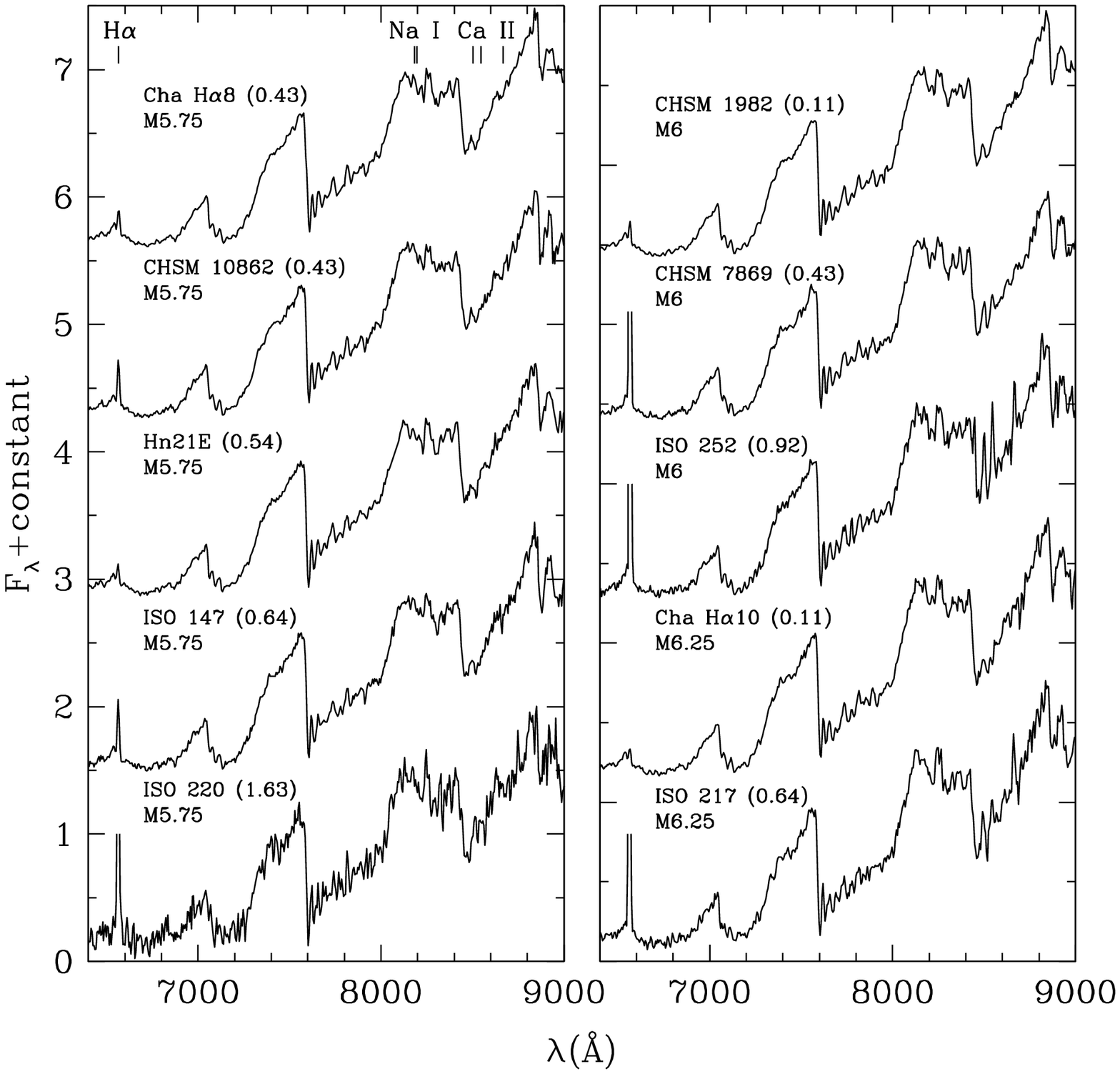}
\caption{Same as Figure~\ref{fig:m1}.}
\label{fig:m7}
\end{figure}

\begin{figure}
\plotone{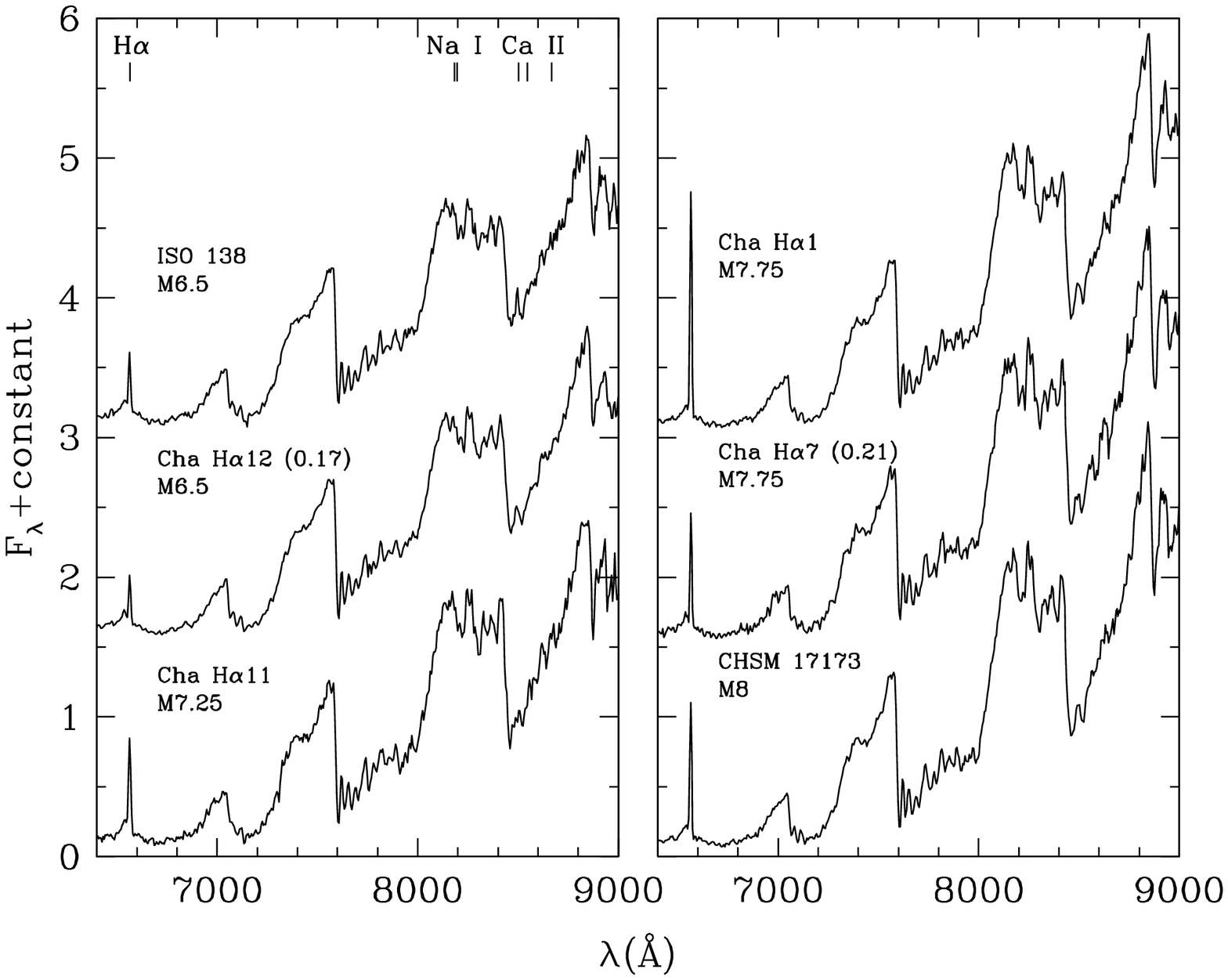}
\caption{Same as Figure~\ref{fig:m1}.}
\label{fig:m8}
\end{figure}

\begin{figure}
\plotone{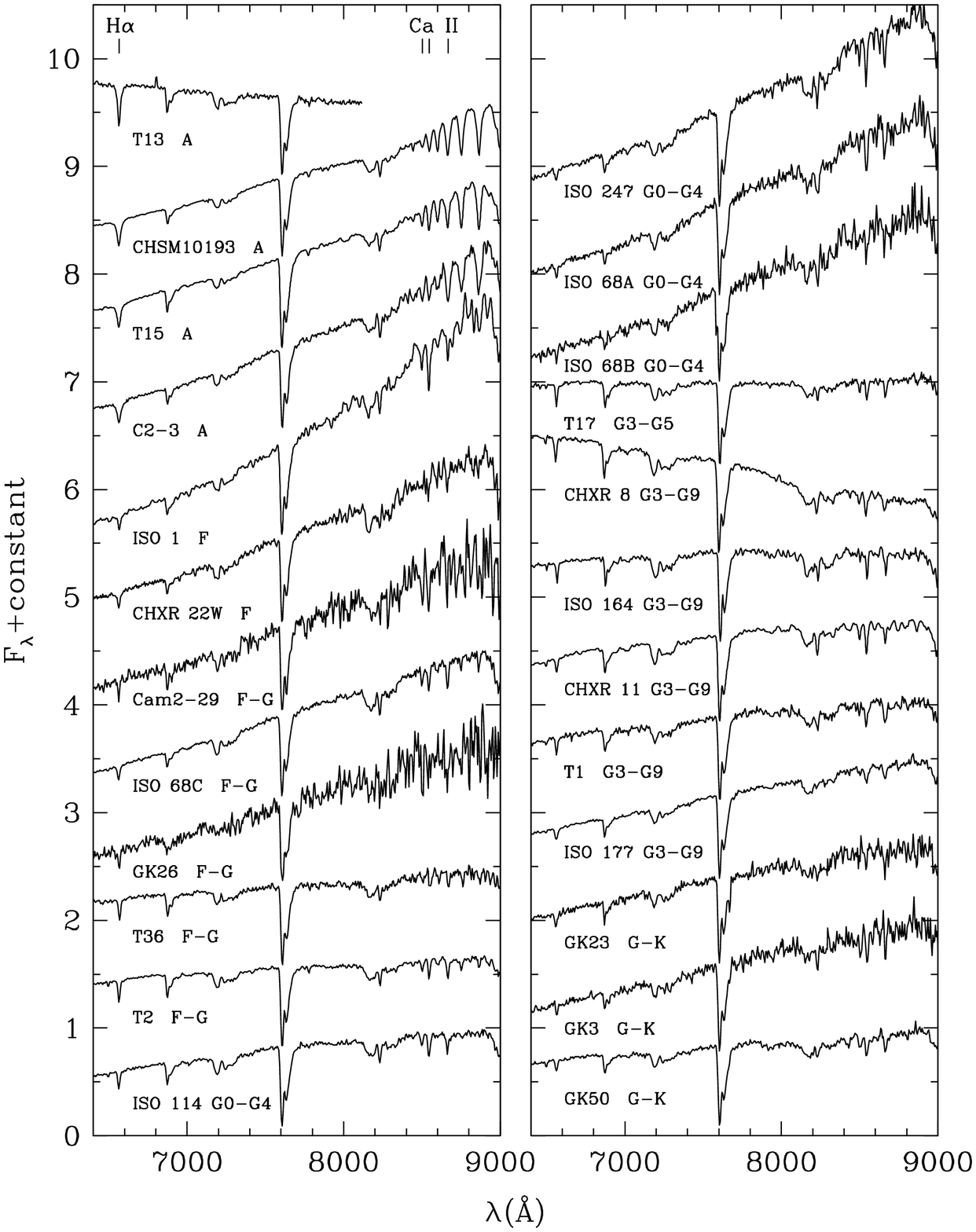}
\caption{Low-resolution spectra of objects toward the Chamaeleon~I star-forming 
region that are classified as early-type background field stars 
(\S~\ref{sec:confirm}).
The data are displayed at a resolution of 13~\AA\ and are normalized at 
7500~\AA.}
\label{fig:fb1}
\end{figure}

\begin{figure}
\plotone{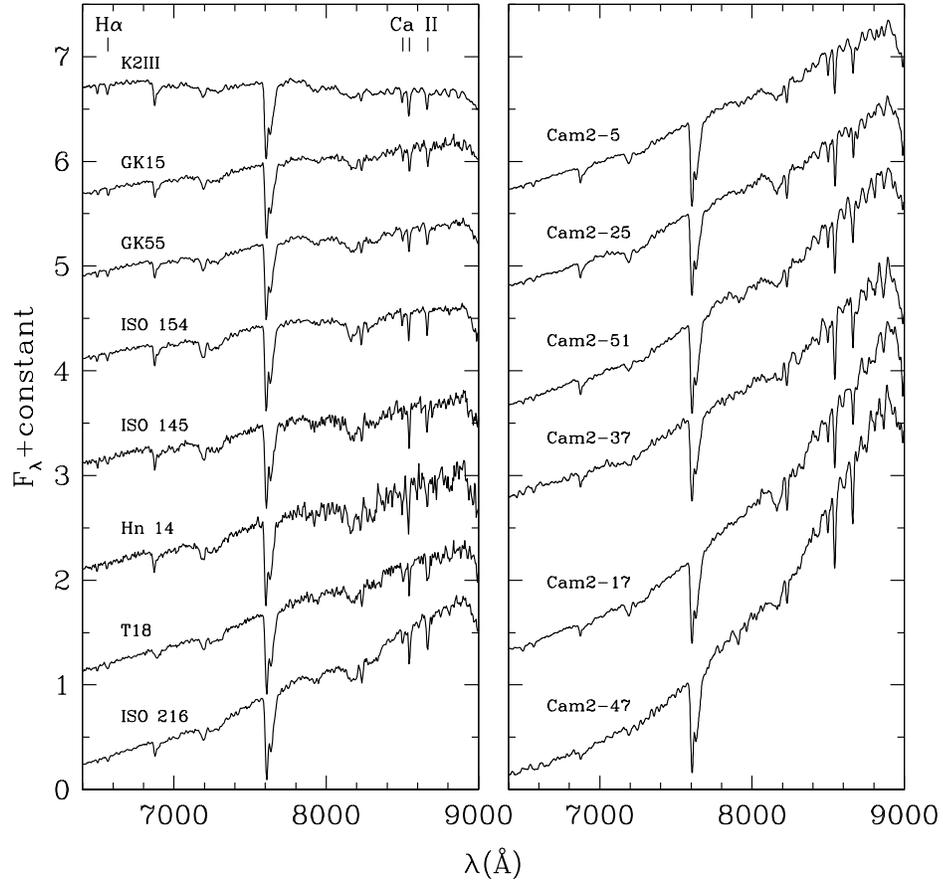}
\caption{Low-resolution spectra of objects toward the Chamaeleon~I star-forming 
region that are classified as K0-K3 background field stars 
(\S~\ref{sec:confirm}).
These stars exhibit the reddening and strong Ca~II absorption that are 
expected of background giants. 
For comparison, a spectral standard star with a type of K2III is shown.
The data are displayed at a resolution of 13~\AA\ and are normalized at 
7500~\AA.}
\label{fig:fb2}
\end{figure}

\begin{figure}
\plotone{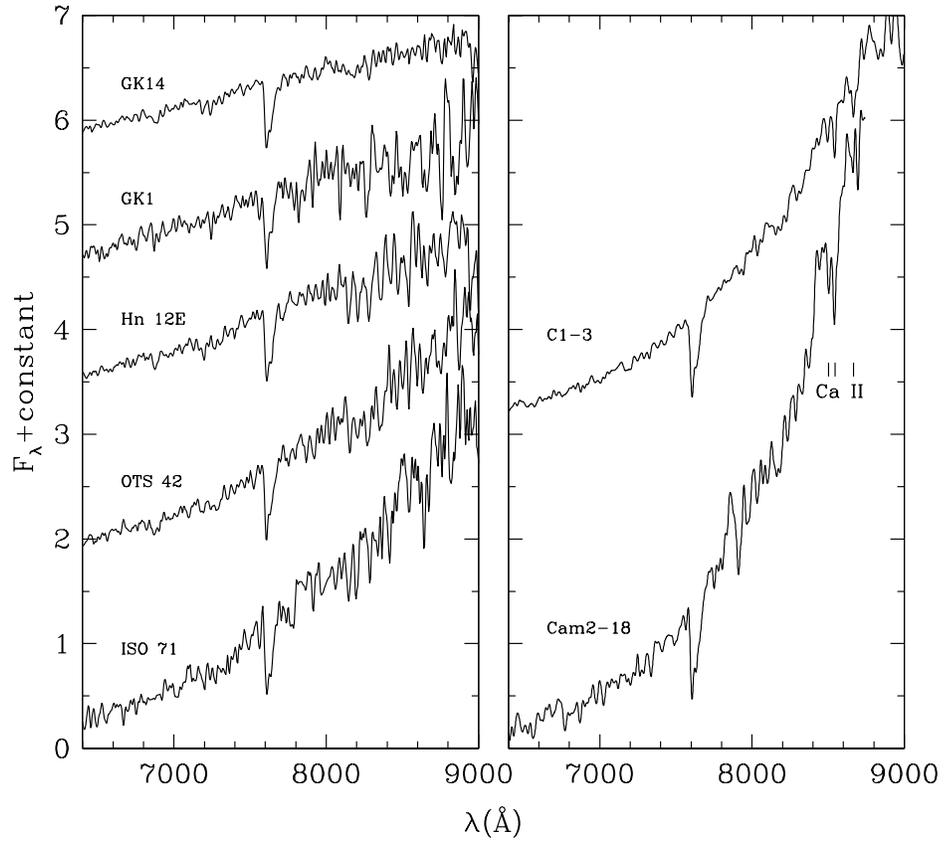}
\caption{Low-resolution spectra of objects toward the Chamaeleon~I star-forming 
region that are classified as background field stars with types earlier than M0
(\S~\ref{sec:confirm}).
The data are displayed at a resolution of 13~\AA\ and are normalized at
7500~\AA.}
\label{fig:fb3}
\end{figure}

The portions of the 600~l~mm$^{-1}$ data that encompass H$\alpha$ and Li are 
shown in Figure~\ref{fig:li}.
The lower-resolution spectra for the targets classified as members of 
Chamaeleon~I in the next section are presented in 
Figs.~\ref{fig:m1}-\ref{fig:m8}. These data have 
been corrected for reddening with the procedure described in \S~\ref{sec:ext}.
The observed spectra of the field stars are shown in 
Figs.~\ref{fig:fb1}-\ref{fig:fb5}. All spectra are available from the author
upon request. 

\section{Classification of Candidate Members}
\label{sec:class}

\subsection{Spectral Types}
\label{sec:sptype}

In this section, I measure spectral types for the 179 candidate members
of Chamaeleon~I that have been observed spectroscopically in this work. 
I also use spectra from \citet{saf03} to classify the four candidates from 
that study that were not included in my spectroscopic sample.

Low-resolution red optical spectra of early-type dwarfs and giants
are characterized by absorption in the Ca~II triplet, H$\alpha$, and a blend 
of transitions near 6500~\AA. At types earlier than mid K, I measured
spectral types by comparing the strengths the latter two features to those 
in the standard dwarfs and giants from \citet{as95} and this work.
Although the Ca~II transitions are sensitive to surface gravity, I did not 
use them to assign luminosity classes because few of the available standard 
giant spectra encompassed these lines.

At types later than mid K, H$\alpha$ disappears and the TiO, CaH, and VO
absorption bands rapidly become stronger with later spectral types. 
The strengths of these bands and the K~I, Na~I, and Ca~II absorption lines 
differ noticeably among dwarfs, giants, and pre-main-sequence objects
\citep{mar96,kir97,luh99}, particularly at later types.
These molecular and atomic features were used to measure spectral types 
and luminosity classes through a comparison to data for dwarfs and giants 
obtained in this work and from \citet{as95}, \citet{kir91}, \citet{hen94}, 
and \citet{kir97}. 
The spectral types of the members of Chamaeleon~I were derived by comparisons
to dwarfs and averages of dwarfs and giants
for types $\leq$M5 and $>$M5, respectively \citep{luh99}.
The typical precision of the optical spectral types is
$\pm0.5$ and $\pm0.25$ for K and M types, respectively, unless noted otherwise.
For the Chamaeleon~I members that exhibit intense emission lines, 
such as T14A, T44, and T42,
strong blue excess continuum emission also may be present,
which would weaken the observed absorption bands and thus imply types that 
are too early. To mitigate the effects of veiling, the spectral types for these
sources are based primarily on the reddest absorption bands ($>$8000~\AA)
where any excess continuum emission should be weakest.

\subsection{Membership}

I now use the spectroscopic data in this work and all available constraints 
from the literature to evaluate the membership status of the potential 
members of Chamaeleon~I that have been presented in previous surveys. 
In the following analysis, I consider all of the candidate members compiled by
\citet{car02}, the 10 new candidates from \citet{car02}, the 
additional candidates from \citet{gk01} and \citet{cam98} that I included 
in my spectroscopic sample (\S~\ref{sec:targets}), and the candidates
observed spectroscopically by \citet{saf03}.
I also examine various sources that have been identified 
as likely field stars in previous studies within a radius of $1\fdg25$ from
$\alpha=11^{\rm h}07^{\rm m}00^{\rm s}$, $\delta=-77\arcdeg10\arcmin00\arcsec$
(J2000), a few which will be classified as members of Chamaeleon~I. 
A total of 288 objects are evaluated.

Stars projected against the Chamaeleon~I clouds can be background stars, 
foreground stars, or young members of the star-forming region.  
Following \citet{luh03b}, I employ several diagnostics to distinguish between
these possibilities, including proper motions, radial velocities, trigonometric
parallaxes, luminosity classifications, reddenings, positions on an H-R 
diagram, and signatures of newly formed stars (emission lines, Li absorption, 
IR excess emission). Although X-ray observations are useful in selecting
candidate young stars, I do not take X-ray emission as evidence of membership 
as it can arise from both active field stars and cluster members.
I next describe each of these diagnostics.
Additional comments on individual sources are in \S~\ref{sec:append}.

\subsubsection{Confirmed Members and Field Stars}
\label{sec:confirm}

The membership of a star can be examined by comparing its proper motion, 
radial velocity, and parallax to the values of the known cluster members.  
If any one of these measurements differs significantly from the cluster mean, 
then it can be confidently identified as a field star. On the other hand, 
consistency with the mean cluster values supports membership but 
does not guarantee it (e.g., T19, \S~\ref{sec:append}).
Because of the low accuracies of most of the published 
measurements of proper motions and parallaxes in Chamaeleon~I, I consider them
in this work only when no other membership constraints are available.

In the previous section, I spectroscopically 
determined the luminosity classes for a large number of sources,
primarily the ones at M spectral types. These classes consist of dwarfs,
giants, and pre-main-sequence objects, and thus differentiate between 
field stars and young cluster members.

Other constraints on membership include a star's reddening and position on the
H-R diagram. If a star exhibits significant reddening in its spectrum or colors 
($A_V\gtrsim1$), then it cannot be a field star in the foreground of the 
cluster. Meanwhile, if a star is clearly not a giant by its spectrum and it 
appears above the main sequence for the distance of Chamaeleon~I, then it 
cannot be a background field star. Objects that are neither foreground stars
nor background stars must be members of the cluster.

\begin{figure}
\plotone{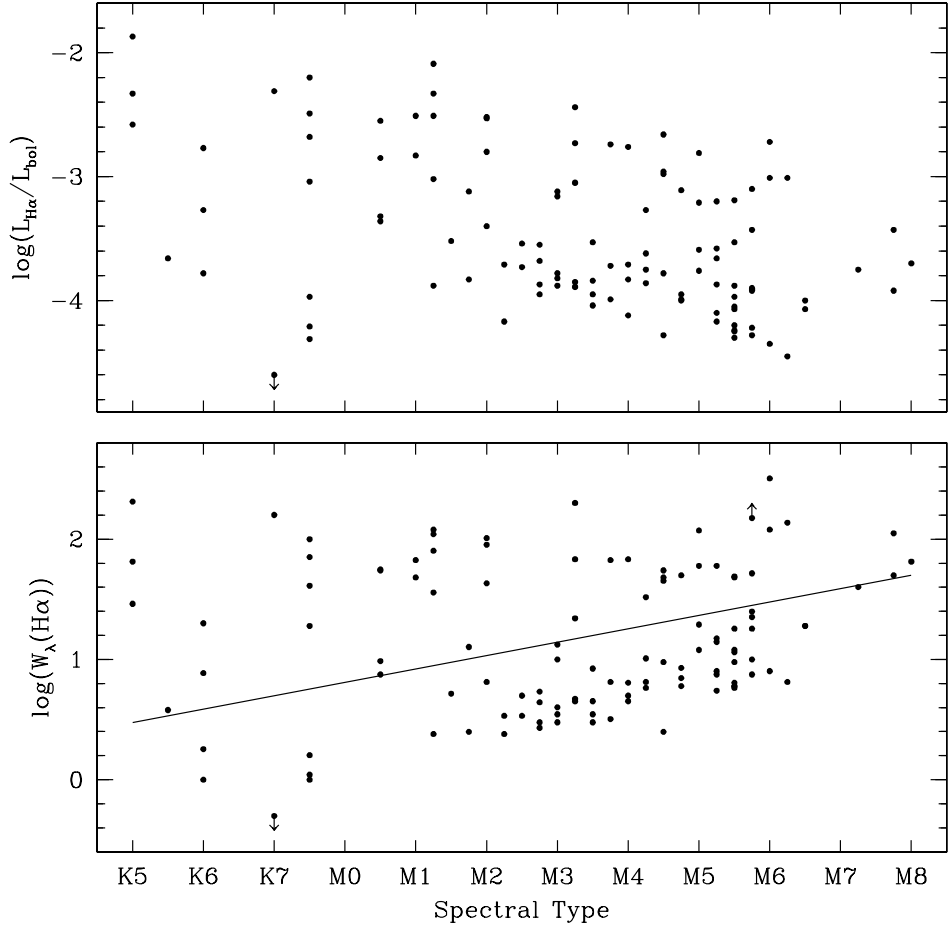}
\caption{
H$\alpha$ emission as a function of spectral type for the members of 
the Chamaeleon~I star-forming region observed in this work (Table~\ref{tab:ha}).
The solid line has been selected as an approximate upper limit of H$\alpha$ 
emission strengths for field dwarfs \citep{giz00,giz02}. 
Therefore, a measurement above this boundary is taken as evidence of youth and
membership in Chamaeleon~I. The endpoints of the line are 3 and 50~\AA\ at
K5 and M8, respectively.}
\label{fig:ha}
\end{figure}

Because young stars are associated with processes such as accretion and 
outflows, members of star-forming regions often exhibit strong emission lines,
examples of which at optical wavelengths include H$\alpha$, He~I, Ca~II, 
and various forbidden transitions.
However, chromospherically active field stars also can produce emission in
H$\alpha$ and other permitted lines. Therefore, emission lines comprise
evidence of youth and membership in a star-forming cluster only if they are
stronger than the lines found in field dwarfs. For the case of H$\alpha$,
I have selected a boundary that approximates the upper limit of emission 
strengths observed in field dwarfs in the diagram of 
log~$W_{\lambda}$(H$\alpha$) versus spectral type in Figure~\ref{fig:ha} 
\citep{sh86,hgr96,del98,giz00,giz02}.
Only data above that line are taken as evidence of membership.
For sources that were observed spectroscopically in this work and that 
are classified as members of 
Chamaeleon~I in this section, Table~\ref{tab:ha} lists the equivalent widths 
and fluxes of H$\alpha$ that I measured from the spectra. The line fluxes
were combined with the bolometric luminosities from \S~\ref{sec:teff} to
arrive at log~$L_{H\alpha}/L_{\rm bol}$ for each source. These measurements
of log~$W_{\lambda}$(H$\alpha$) and log~$L_{H\alpha}/L_{\rm bol}$ are 
plotted in Figure~\ref{fig:ha}. 
The adopted threshold for identifying a young star via H$\alpha$ emission 
is fairly conservative; a field dwarf could be mistaken as a young star 
only if it exhibited an unusually intense flare. 
In fact, for all of the sources above the boundary in Figure~\ref{fig:ha},
membership is supported by at least one other diagnostic in this section.
In addition to H$\alpha$, detections of emission in forbidden optical
transitions or IR lines were taken as evidence of youth as well.

After the formation of a low-mass star, convection and 
nuclear burning steadily deplete Li at its stellar surface over time.
As a result, Li absorption is stronger in young stars than in 
their older counterparts in the field. The change in Li strength with age
is particularly large at M spectral types, where the equivalent widths 
vary from 0.45-0.8~\AA\ at 1~Myr in Taurus \citep{bas91,mar94}
to nearly zero at 30~Myr in IC~2602 \citep{ran97}.
Meanwhile, for stars at earlier types, the initial Li absorption is 
weaker and the species is not fully depleted even at 100~Myr, resulting in 
a smaller change in observed Li strengths with time and reducing the 
effectiveness of Li as a youth diagnostic at types earlier than K0.
Therefore, as with emission lines, Li absorption is evidence of 
membership only if the equivalent width is greater than the maximum values
for the youngest field stars ($\gtrsim10$~Myr) at a given spectral type.
In this comparison, young field stars can be represented by members of the 
Pleiades and IC~2602 open clusters \citep{sta97,ran97}. The strength of Li
can be significantly overestimated in low-resolution data because of
blending with nearby Fe lines \citep{bas91,cov97,bri97}.
The highest resolution of 3~\AA\ obtained 
in this work should provide sufficiently accurate measurements \citep{luh01},
particularly at the late-K and M types of most of the targets observed at 
that setting. These spectra are shown in Figure~\ref{fig:li}, where the
young members of Chamaeleon~I are readily distinguished from old field 
dwarfs and giants via Li. The Li measurements for the objects observed
at high resolution and classified as members are provided in Table~\ref{tab:li}.

\begin{figure}
\plotone{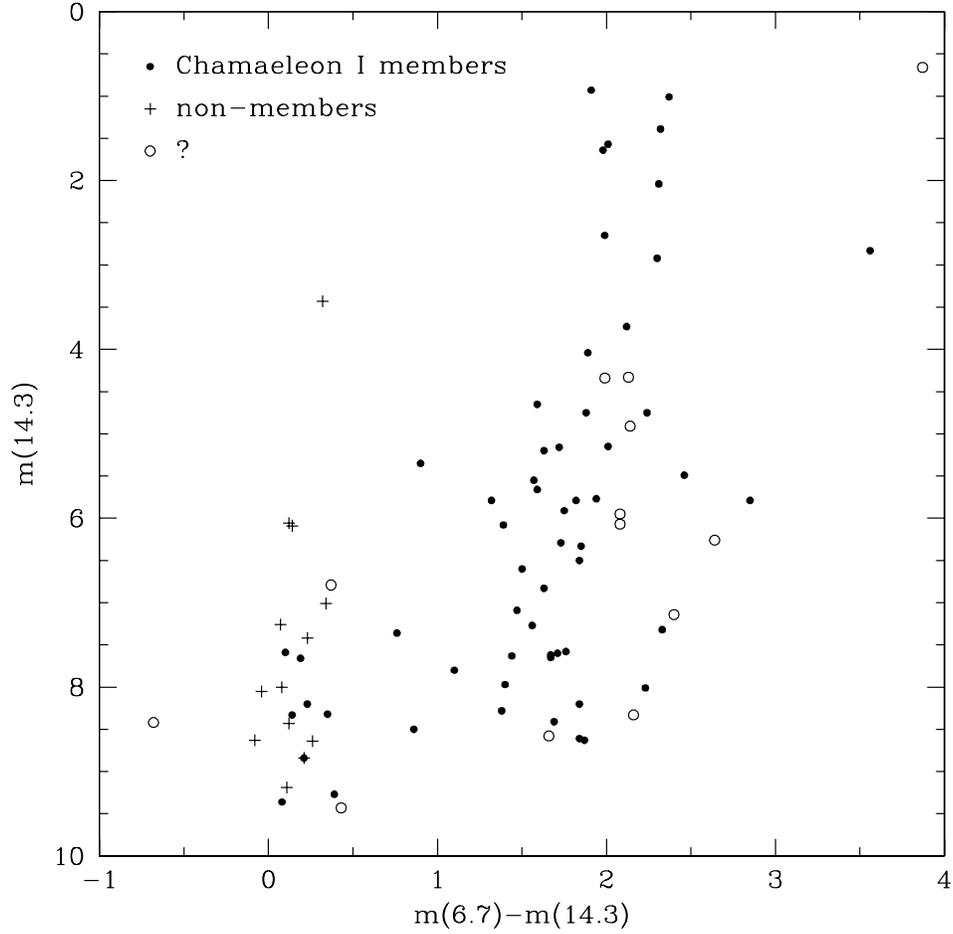}
\caption{
$m(6.7)-m(14.3)$ versus $m(14.3)$ for objects that are 
classified as members ({\it points}) and as non-members ({\it plusses}) 
of the Chamaeleon~I star-forming region through the diagnostics in 
\S~\ref{sec:confirm}, with the exception of IR excesses. 
For sources whose membership could not be determined in that way 
({\it circles}), $m(6.7)-m(14.3)>1$ is taken as evidence of youth and 
membership in the remainder of this work since field stars do not exhibit
these colors.
A reddening vector in this diagram would be roughly vertical.
These measurements are from \citet{per00}.}
\label{fig:iso1}
\end{figure}

As with emission lines and Li absorption, emission from cool circumstellar 
material in excess above that from the photosphere of the central star is 
a signature of youth that can establish the membership of an object. 
With longer IR wavelengths, these excesses become larger and produce colors
that are distinctive from those of field stars. 
For instance, Figure~\ref{fig:iso1} presents 6.7 and 14.3~\micron\ {\it ISO} 
photometry \citep{per00} for stars that are determined to be members and 
field stars by the previous diagnostics in this section and for the remaining 
stars that lack the data for those diagnostics. In a diagram of this kind, 
field stars are well-separated from the young stars that exhibit IR excesses 
\citep{nor96,olo99,per00}. Therefore, a color of $m(6.7)-m(14.3)>1$ is taken
as evidence of membership in this work. 
Meanwhile, as shown in the next section, most young stars are not unambiguously 
distinguished from field stars through colors at shorter IR wavelengths.
In this work, only excesses measured at $\lambda\geq10$~\micron\ are used 
to establish youth.

\begin{figure}
\plotone{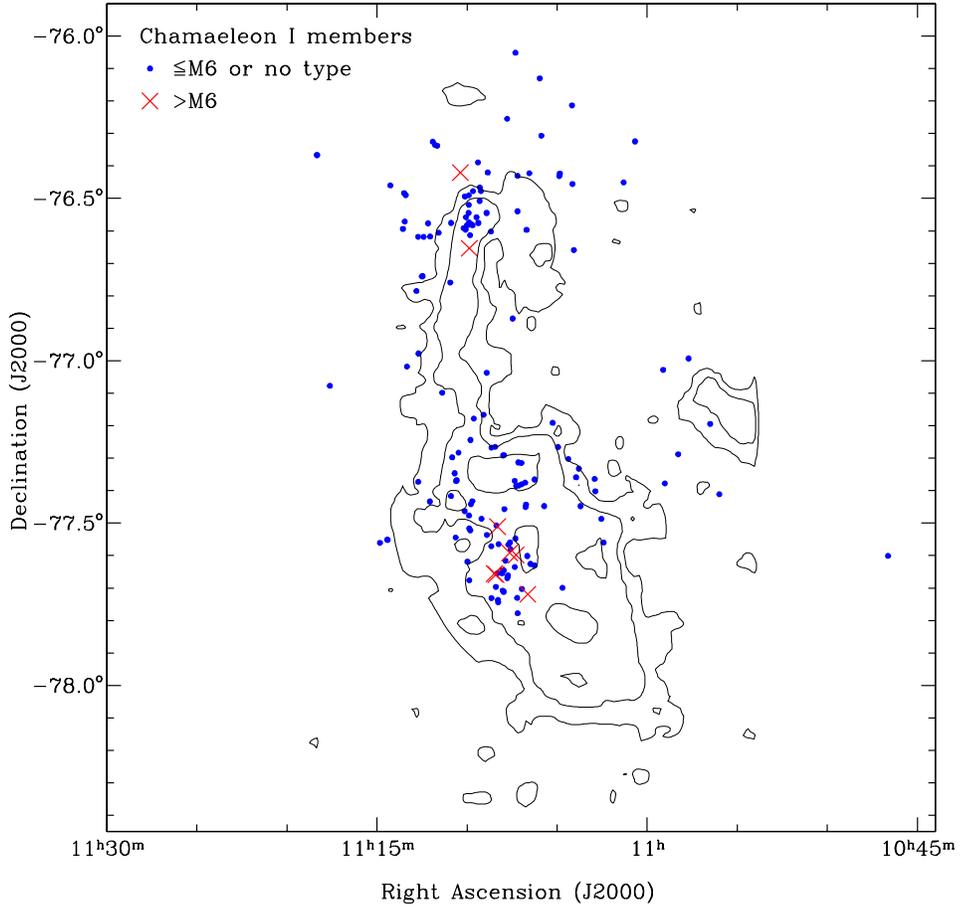}
\caption{
Spatial distribution of the 158 objects that are classified as members of the 
Chamaeleon~I star-forming region (\S~\ref{sec:confirm}). The eight members with 
spectral types later than M6 are likely to be brown dwarfs according to the
H-R diagram and evolutionary models in Figure~\ref{fig:hr}. 
The contours represent the extinction map of \citet{cam97} at intervals 
of $A_J=0.5$, 1, and 2.}
\label{fig:map}
\end{figure}

I have applied the above membership diagnostics to all relevant data from 
the literature for the 288 sources considered here. For the 158 objects 
classified as members of Chamaeleon~I through this analysis, astrometry and
photometry are listed in Table~\ref{tab:mem1} and spectral types and membership
evidence are listed in Table~\ref{tab:mem2}. T33A and B are combined into one
entry in these lists but are otherwise counted separately in this work.
The membership for 41 of these sources is based on data collected in
this work alone.
Tables~\ref{tab:fore} and \ref{tab:back} contain similar data for the 13
foreground stars and 91 background sources, respectively. 
The membership of the 26 stars listed in Table~\ref{tab:unc} cannot be
determined with the available data. 
I have adopted $J$, $H$, and $K_s$ measurements from the Two-Micron
All-Sky Survey (2MASS) Point Source Catalog and $i$ from the 
Deep Near-IR Survey of the Southern Sky (DENIS) Second Release. 
When multiple $i$ measurements were listed for a given star in
the latter database, the average of those value has been adopted. 
For 30 of the stars that do not have $i$ in the DENIS Second Release,
I have adopted the preliminary DENIS data from \citet{cam98}.
Most of these stars are within two areas that are not in the DENIS Second 
Release, which are defined approximately by 
$\alpha=11^{\rm h}03^{\rm m}15\fs3$ to 
$11^{\rm h}04^{\rm m}51\fs6$ and
$\alpha=11^{\rm h}11^{\rm m}16\fs8$ to 
$11^{\rm h}12^{\rm m}55\fs8$ (J2000).
In Figure~\ref{fig:map}, the positions of the members of Chamaeleon~I are 
plotted with the extinction map from \citet{cam97}.

\subsubsection{Candidate Members and Field Stars}
\label{sec:cand}

Definitive membership classifications are not possible for the sources in 
Table~\ref{tab:unc}. The following methods can identify some of these objects
as either candidate members or candidate field stars. 

\begin{figure}
\plotone{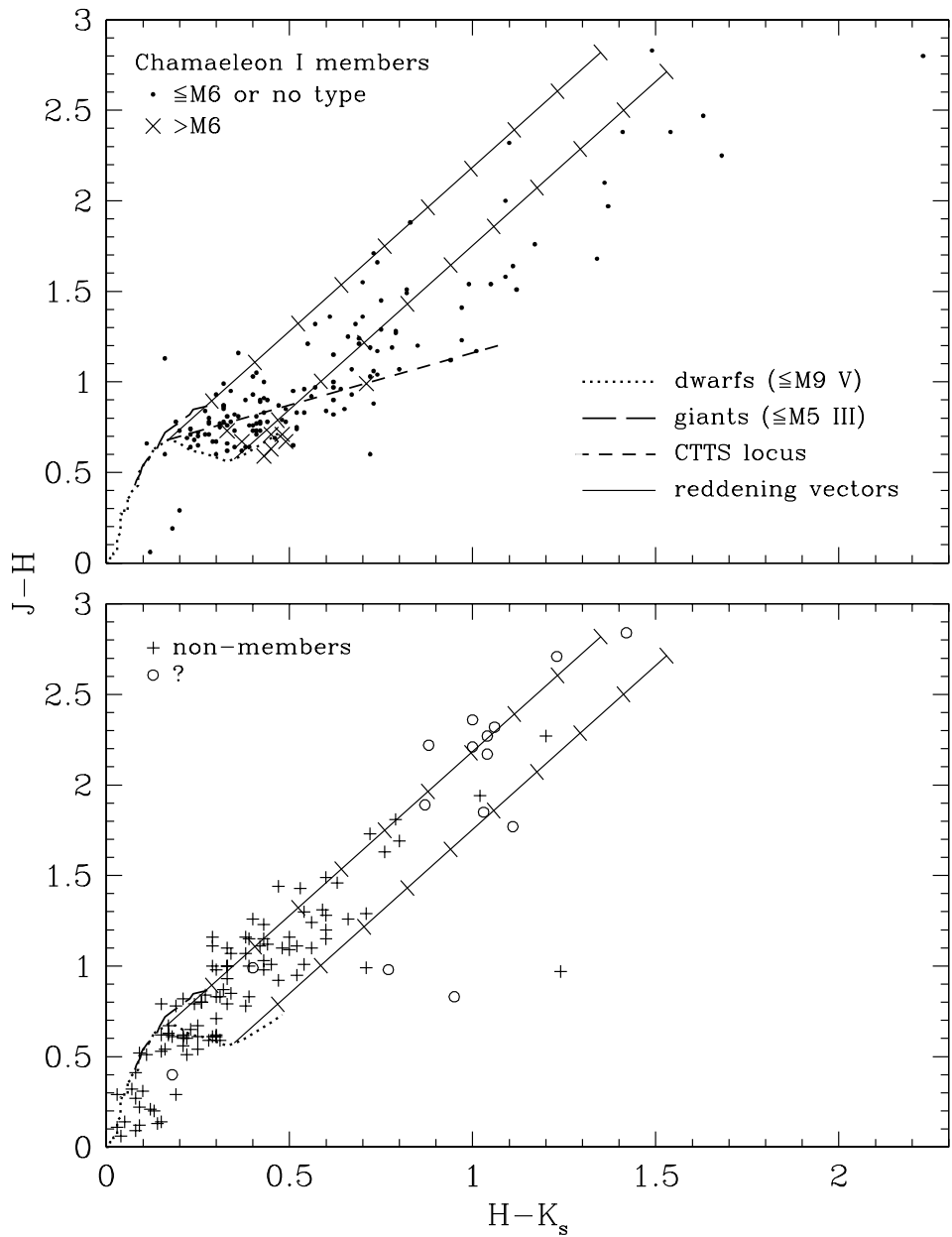}
\caption{
$H-K_s$ versus $J-H$ for objects that are classified as members 
({\it points and crosses}) and as non-members ({\it plusses}) of the 
Chamaeleon~I star-forming region (\S~\ref{sec:confirm}). 
The eight members with spectral types later than M6 are likely to be brown 
dwarfs (Figure~\ref{fig:hr}). 
Sequences for typical field dwarfs ({\it dotted line};
$\leq$M9V) and giants ({\it long dashed line}; $\leq$M5~III) are plotted
with reddening vectors originating at M0~V ({\it left solid line}) and M6.5~V
({\it right solid line}), which are marked at intervals of $A_V=2$.
The upper panel also includes the locus of M0 classical T~Tauri stars in 
Taurus as measured by \citet{mey97} ({\it short dashed line}).
Three of the sources whose membership could not be determined 
({\it circles}; Table~\ref{tab:unc}) are redder in $H-K_s$ than expected
for reddened field stars earlier than M6.5 and therefore may exhibit IR 
excess emission that would indicate youth and membership in Chamaeleon~I.
These measurements are from 2MASS.}
\label{fig:jhhk}
\end{figure}

Although the contrast of cool circumstellar material against a stellar 
photosphere is lower at shorter wavelengths, excess emission from the former
is still detectable in colors at 1-10~\micron.
In Figure~\ref{fig:iso2}, I plot the objects classified as members and as
field stars in the previous section in a diagram of $K_s-m(6.7)$ versus
$J-K_s$. The overlap between young stars and field stars is greater than in
the diagram based on longer-wavelength photometry in Figure~\ref{fig:iso1},
but this may be partially due to erroneous measurements. For instance, the
two reddest field stars in $K_s-m(6.7)$, ISO~145 and CHXR~22W, are only 
$6\farcs2$ and $10\farcs6$ from stars that are brighter at $K_s$ by 
1.9 and 2.5 magnitudes, respectively. The 6.7~\micron\ measurements 
for ISO~145 and CHXR~22W may have been contaminated by emission from these
nearby stars, which would result in false excesses at 6.7~\micron\ relative
to the 2MASS bands. At any rate, upon inspection of Figure~\ref{fig:iso2},
I find that one of the unclassified stars, ISO~206, may have a modest
excess in $K_s-m(6.7)$ and thus is a potential member of Chamaeleon~I.
A similar exercise is performed with the diagram of $H-K_s$ versus $J-H$
in Figure~\ref{fig:jhhk}. 
I plot the standard dwarf colors from \citet{kh95} ($\leq$M0) and
from the young disk populations in \citet{leg92} ($>$M0). The former 
are transformed from the Johnson-Glass photometric system to the CIT 
system \citep{bb88}, which is equivalent to the 2MASS system at
a level of $<0.1$~mag \citep{car00}. 
Three of the objects with undetermined membership,
OTS~7, 32, and 44, are on the right side of the reddening vector for M6.5 
dwarfs and thus appear to have IR excesses that are indicative of youth. 
Although the 2MASS data that are used for these three sources have large
uncertainties, they do support the presence of excesses originally reported 
by \citet{ots99}. Young stars are not the only sources that exhibit colors
redward of the reddening band of main sequence stars, as illustrated in 
Figure~\ref{fig:jhhk} by CHSM~11564 (Seyfert galaxy) and IRAS~11248-7653 
(Mira variable).

\begin{figure}
\plotone{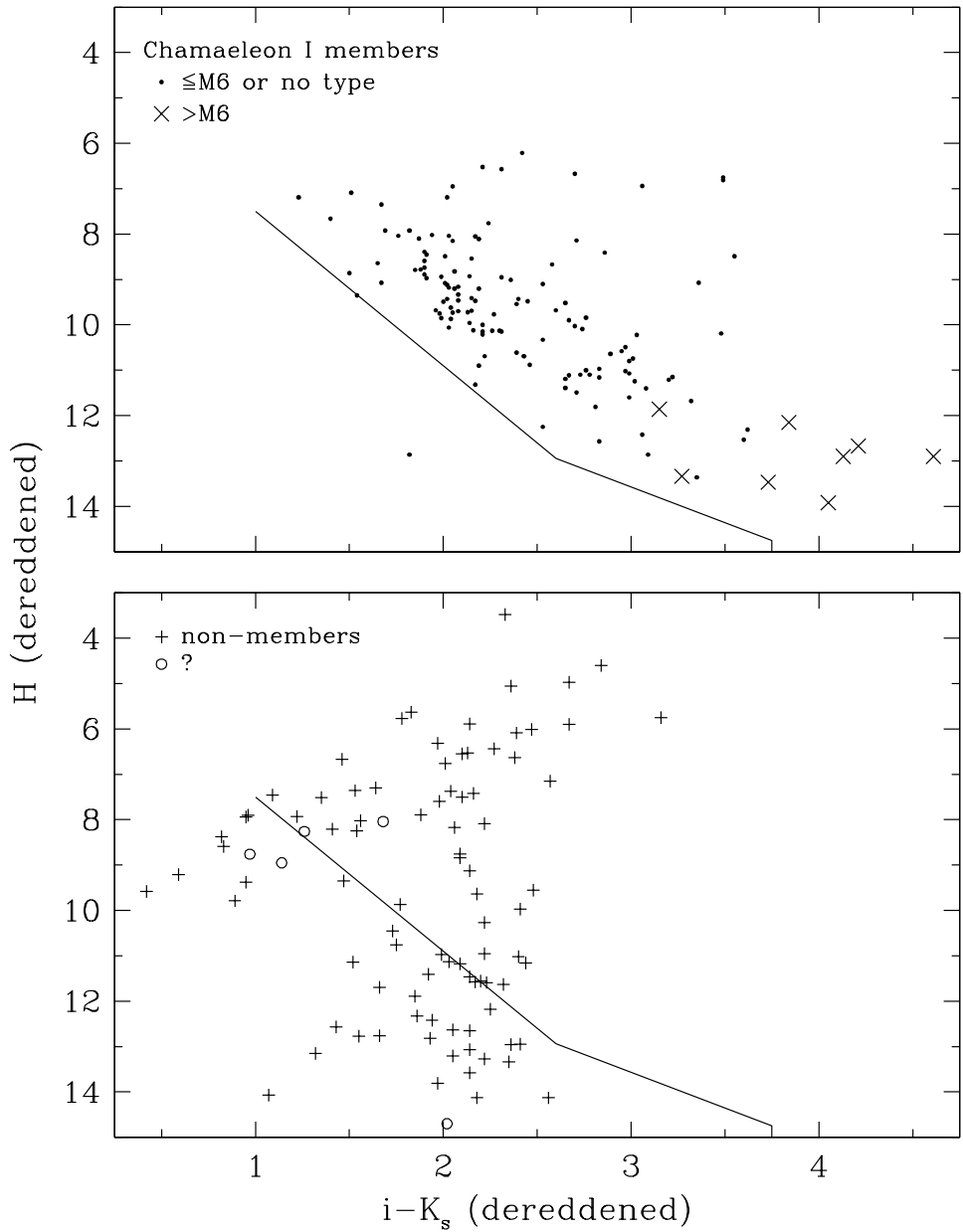}
\caption{
Extinction-corrected color-magnitude diagrams for objects that are classified 
as members ({\it points and crosses}) and as non-members ({\it plusses}) of the 
Chamaeleon~I star-forming region (\S~\ref{sec:confirm}). The eight members with 
spectral types later than M6 are likely to be brown dwarfs 
(Figure~\ref{fig:hr}).
The solid boundary was designed to follow the lower envelope of the sequence of
members. Among the sources whose membership could not be determined 
({\it circles}; Table~\ref{tab:unc}), stars that are above and below this
boundary are candidate members and likely field stars, respectively.
These measurements are from DENIS ($i$) and 2MASS ($H$, $K_s$).}
\label{fig:ik}
\end{figure}

A color-magnitude diagram composed of optical and near-IR photometry is
another useful tool in distinguishing candidate members of a young cluster
from likely field stars. 
Following \citet{luh03a} and \citet{luh03b}, I select extinction-corrected 
$i-K_s$ and $H$ for the color-magnitude diagram of the objects 
considered here. In those previous studies, the extinction for a given star
was computed by dereddening its position in a diagram of $J-H$ versus 
$I-K_s$ until it intersected the sequence defined by field M dwarfs. 
However, if a star exhibits a very large $K$-band excess, this procedure 
will overestimate its extinction. 
Therefore, in this work, if $H-K_s$ is more than 0.1~mag greater than the 
reddening vector for M9~V, then the extinction is instead computed from 
$J-H$ assuming an intrinsic color of 0.9, which is a typical intrinsic 
value for a young star with a large excess \citep{mey97}. 
This dereddening method used here and by \citet{luh03a} and 
\citet{luh03b} requires no knowledge of spectral types or membership
and can be applied uniformly to photometry to better separate cluster
members from field stars in a color-magnitude diagram. More accurate 
extinctions for the confirmed cluster members will be estimated in 
\S~\ref{sec:ext} by incorporating the spectral type measurements. 
Figure~\ref{fig:ik} contains the extinction-corrected diagram of
$i-K_s$ versus $H$ for members, field stars, and unclassified sources.
The members of Chamaeleon~I form a well-defined sequence, the lower envelope
of which has been outlined by the boundary plotted in Figure~\ref{fig:ik}.
ISO~225 is the only known member that appears below this boundary. It is
anomalously faint for its spectral type in an H-R diagram as well
(\S~\ref{sec:hr}). Three of the six unclassified sources for which 
$i$, $H$, and $K_s$ are available, Cam2-12, Cam2-14, and OTS~61, are 
below the sequence of members and therefore are probably field stars.
Cam2-35/CHSM~17388, is outside the limits of 
Figure~\ref{fig:ik} and is below the sequence as well, but its optical
and IR measurements are uncertain because of nebulosity. Cam2-18 is a 
marginal candidate by its position just above the boundary in 
Figure~\ref{fig:ik}. C1-14 is the other star above the boundary 
and thus is a possible member (\S~\ref{sec:append}).

\section{Implications for Previous Surveys}
\label{sec:imp}

In the previous section, I measured spectral types for 179 objects toward
Chamaeleon~I. I then used these spectra and published data to
identify 158 members of Chamaeleon~I, 104 non-members, and 26 sources that
could not be classified in a sample of 288 potential members from previous
studies. Among the members, 137, 8, and 13 objects have spectral types that
are $\leq$M6, $>$M6, and not measured, respectively. In this section, I examine 
the implications of this census for the results from recent searches for 
members of Chamaeleon~I.

\citet{gp02} obtained low-resolution near-IR spectra of 13 of the 34 candidate
members of Chamaeleon~I presented by \citet{per00} and measured
spectral types from the steam absorption bands in these data. For some of 
these candidates, the spectra also provided evidence of youth and membership 
in the form of emission lines and possible continuum veiling. 
One of these objects, ISO~98/Cha~H$\alpha$~9, had been previously classified
and confirmed as a member by \citet{com00}. The optical spectral types from
the previous section agree fairly well with the classifications of 
\citet{gp02} with a few exceptions, most notably ISO~154. The presence
of absorption in H$\alpha$, the strength of Ca~II, and the lack of TiO 
absorption in the optical spectrum of this object in Figure~\ref{fig:fb2}
clearly demonstrate that it is a background K giant rather than an M0 star 
as reported by \citet{gp02}.

In a similar study to that of \citet{gp02}, \citet{gm03} recently presented
near-IR spectra of 46 candidate members of Chamaeleon~I that were found in 
various previous surveys. No useful information was derived for 15 of these 
objects because of insufficient signal-to-noise. Two stars exhibited strong
absorption in Br$\gamma$ and were classified as earlier than M0. 
For the remaining 29 sources, \citet{gm03} estimated spectral types from the
steam absorption bands. They then proceeded to treat these 
objects as members of Chamaeleon~I, placing them on the H-R diagram and
estimating their masses and ages. However, the only evidence of membership
in Chamaeleon~I that was presented by \citet{gm03} consisted of emission line
detections in two stars. Otherwise, they failed to demonstrate that
the other targets were members rather than field stars
through their spectroscopy or any other means. Indeed, on the surface,
the simple fact that no field stars were reported among 13 and 29 candidates 
observed by \citet{gp02} and \citet{gm03} is surprising.
Based on the optical spectra in the previous section, I find that 12 of 
the 29 targets from the latter study are in fact background stars, most 
of which are clearly field giants. These results are consistent with the
spectra in \citet{gm03}, which I have been able to classify through visual 
inspection since giants are readily distinguished from young stars with
$K$-band spectra (e.g., \citet{lr99}). In addition, after inspection of
the data in \citet{gm03} for sources not in my optical sample, I suggest that
Cam2-12, 14, and 33, and probably 34 and 36, are background giants as well.
This high contamination by background stars ($\gtrsim60$\%) is the source of 
the low frequency of emission lines among the candidates in \citet{gm03} (2/29) 
relative to previously known members (20/63) that was noted in that study.
Among the remaining 12 objects from \citet{gm03}, two sources lack independent
constraints on their membership (Cam-8, 43) and 10 sources are classified 
as members in this work. 
For one of these members, GK18/CHSM~10862, \citet{gm03} measured a spectral
type of M6.5 and estimated a mass of 0.04~$M_\odot$, while I arrive at 
M5.75 and a mass of 0.1~$M_\odot$ through the H-R diagram and evolutionary 
models in \S~\ref{sec:hr}. GK18 is one of the candidate members of 
Chamaeleon~I that were identified through near-IR excess emission by 
\citet{gk01}. Eight additional candidates from that sample were observed
with spectroscopy in this work, all of which are field stars, which is 
consistent with the fact that only a small fraction of those candidates, 
including GK18, were confirmed to have excesses by \citet{car02}.

Spectroscopy at optical wavelengths also has been applied to the newest
candidate members of Chamaeleon~I. \citet{saf03} obtained low-resolution
optical spectra for eight candidates and investigated their membership,
primarily through H$\alpha$.
They identified probable cluster members and field stars by the
presence and absence of H$\alpha$ emission, respectively.
However, these criteria are flawed as emission in H$\alpha$ is moderately 
strong in some active field dwarfs \citep{giz02} and weak in some young stars. 
Instead, as discussed in \S~\ref{sec:class}, H$\alpha$ can serve as a
membership diagnostic only if it is sufficiently strong, in which case it 
provides evidence of youth. Otherwise, H$\alpha$ at lower levels is not 
useful in discriminating between field stars and young stars.
Indeed, \citet{saf03} suggested that ISO~237 was a possible field star because 
of weak H$\alpha$, whereas it is clearly a member by the diagnostics from
\S~\ref{sec:class}.
\citet{saf03} also used Na~I to distinguish between young stars and field 
dwarfs among some of the candidates in their sample. Their membership 
classifications based on this feature agree with those in this work for the
most part. The one exception is GK53, whose Na~I absorption is intermediate 
between that of a dwarf and a young star according to \citet{saf03}, but 
is well-matched to the value for a dwarf in my classification. 
GK53 also appears below the sequence of known members in Figure~\ref{fig:ik}
and is near the main sequence when placed on an H-R diagram for the distance
of Chamaeleon~I, which further indicates that it is a field star. It probably
is close to the opposite side of the Chamaeleon~I cloud given its position 
on the H-R diagram and the presence of reddening in its colors.
Finally, \citet{saf03} measured a spectral type
for SGR1 through only a weak detection of TiO at 7100~\AA, which I find to 
be insufficient for a spectral type measurement upon reclassification 
of their data (\S~\ref{sec:class}).

\citet{cam98} used $iJK_s$ photometry from the DENIS survey to search
for candidate young stars in a $1\fdg5\times3\arcdeg$ field toward Chamaeleon~I.
They began this work by compiling a list of 126 previously known members of 
the cluster. Based on the census in the previous section, that list actually 
consists of 103 members, 21 field stars, and one object whose membership status 
is undetermined (Hn10W). The final source, Cam2-67/C9-1, was not included in
my census because it is not a point source in the 2MASS data and instead
appears to be an inhomogeneity in reflection nebulosity \citep{car02}. 
Among the 54 new candidate members presented by \citet{cam98}, I find
10 members, 13 field stars, and nine unclassified sources. Five of the latter
objects were identified earlier in this section as likely background giants 
by examination of the spectra from \citet{gm03} (Cam2-12, 14, 33, 34, 36). 
One of the candidates from \citet{cam98}, Cam2-30, does not appear in either
the DENIS Second Release or the 2MASS catalog. 
The remaining 21 candidates were not considered in the classifications in the
previous section. Fourteen of these sources were detected at $i$ 
and can be placed on a diagram of $i-K_s$ versus $H$ to assess 
candidacy for membership in Chamaeleon~I, as done in \S~\ref{sec:cand}.
Through this exercise, I find nine likely field stars and five candidate 
members (Cam2-6, 26, 27, 28, 52). No constraints on membership are available 
for the seven objects that were detected only at near-IR wavelengths.

By combining 6.7 and 14.3~\micron\ {\it ISO} photometry with near-IR data 
from DENIS, \citet{per00} searched for new members of Chamaeleon~I through
the presence of IR excess emission. They compiled a list of 48 previously 
known members that exhibited IR excess emission. One of these objects, 
CHXR~22W, is classified as a field star in this work. For two other
stars, C1-25 and C1-2, conclusive evidence of membership was not available
until the detections of excesses at 14.3~\micron\ by \citet{per00}. 
The remaining 45 stars are indeed members of Chamaeleon~I according to the 
census presented here. Next, \citet{per00} claimed to discover 34 new young
stars in Chamaeleon~I. Eighteen of these stars have been confirmed as members 
through diagnostics independent from the IR excesses in that study. 
All of the objects with detections at 14.3~\micron\ that were observed
spectroscopically have been confirmed as members (Figure~\ref{fig:iso1}).
Meanwhile, I find that 10 of the 34 new sources from \citet{per00}
are field stars rather than members of Chamaeleon~I, all of which lack 
detections at 14.3~\micron\ and were originally identified as young stars 
by \citet{per00} only because of putative excesses at 6.7\micron. 
For the remaining six new stars
from \citet{per00}, the IR photometric data from that study represent the only
constraints on membership. Five of these sources have large excesses
at 14.3~\micron\ and therefore were taken to be members in \S~\ref{sec:confirm}.
The final object, ISO~206, was not detected at 14.3~\micron\ and is
a candidate member (\S~\ref{sec:cand}).

\citet{car02} identified 10 candidate members of Chamaeleon~I through the
presence of either near-IR excess emission or variability. 
Nine of these sources have been observed spectroscopically in this work.
Seven candidates are confirmed as members while the others are classified
as a background early-type star and a Seyfert galaxy (\S~\ref{sec:append}). 
Evidence of youth and membership for one of these members, GK18/CHSM~10862, 
also was presented by \citet{gm03}. 
When \citet{car02} combined the observed IR photometry of their candidates
with a theoretical mass-luminosity relation for an age of 2~Myr, they
derived masses that were below the hydrogen burning limit for the seven
confirmed members.
However, when placed on the H-R diagram (\S~\ref{sec:hr}), six of these 
objects exhibit ages greater than 2~Myr and thus higher masses 
($\gtrsim0.1$~$M_\odot$). One of these sources is anomalously faint for 
its spectral type, which indicates that it may be observed in scattered light
(\S~\ref{sec:hr}). 
The H-R diagram does imply a substellar mass of $\sim0.03$~$M_\odot$
for the seventh new member, CHSM~17173, which is the coolest known member of 
Chamaeleon~I at a spectral type of M8. 

\section{Properties of Known Members}
\label{sec:prop}

For the 144 known members of Chamaeleon~I that have spectral type measurements
and resolved photometry (i.e., excluding T33B),
I estimate extinctions, effective temperatures, and bolometric luminosities. 
I then place these sources on the H-R diagram and use theoretical evolutionary
models to examine their masses and ages.

\subsection{Extinctions}
\label{sec:ext}

The amount of extinction toward a young star can be estimated from the
reddening of either its spectrum or its broad-band colors. To ensure that the
color excesses reflect only the effect of extinction, contamination from short 
and long wavelength excess emission arising from accretion and disks is 
minimized by considering a wavelength range of $\sim0.6$-1.6~\micron, 
corresponding to photometric bands between $R$ and $H$.
Therefore, given the data that are available for the members of Chamaeleon~I,
I select the optical spectra from this work and the $J-H$ and $H-K_s$ colors 
from 2MASS for measuring extinctions. In the following analysis, I estimate 
color excesses from these two sets of data, compare the excesses to determine
the appropriate reddening relations, and compute extinctions for the cluster 
members from these excesses and relations.

For each member, a color excess in $J-H$, denoted as $E(J-H)$, is estimated by 
dereddening the $J-H$ and $H-K_s$ colors to the locus observed for classical 
T~Tauri stars (CTTS) by \citet{mey97}. The locus is constructed such
that the origin coincides with the dwarf colors for the star's spectral type
and the slope is that measured by \citet{mey97} for M0 stars, which they
predicted to remain relatively constant with spectral type. 

To measure the reddening-induced color excesses exhibited by the 
optical spectra, I require estimates of the unreddened spectra 
as a function of spectral type for young objects in Chamaeleon~I. 
Because of the large number of cluster members that were observed, 
I am able to derive these intrinsic spectra in the following manner. 
I select the bluest observed spectrum at each spectral type, to which I then 
apply the correction for reddening (typically $A_V=0$-0.5) that is necessary 
to reproduce the change in spectral slopes as a function of spectral type 
that is exhibited by the standard dwarfs and giants from \S~\ref{sec:class}. 
In other words, although the intrinsic spectral slope of a young source 
may differ from that of a dwarf or a giant at that spectral type, I assume
that the variation of the slope with spectral type is the same. 
The spectra derived in this way are valid representations of the intrinsic 
spectra only if at least a few of the members have approximately no extinction 
($A_V<0.5$). This assumption is tested in the upcoming 
comparison of the color excesses measured from the optical and near-IR data.
For each cluster member in my spectroscopic sample, I then deredden its 
spectrum to match the intrinsic template spectrum at the spectral type in 
question. This reddening is quantified by the color excess between 
0.6 and 0.9~\micron, denoted as $E(0.6-0.9)$.

\begin{figure}
\plotone{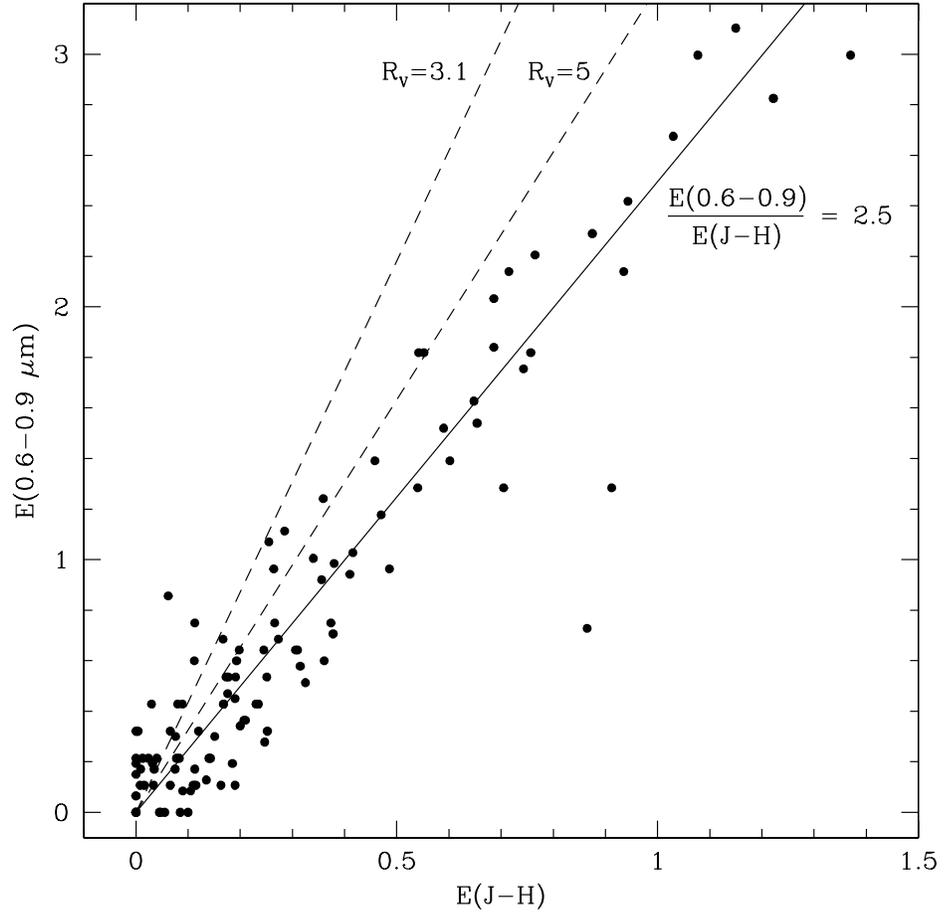}
\caption{
Color excesses estimated from near-IR colors and optical spectra for 
members of the Chamaeleon~I star-forming region.
Relations between these excesses derived from the extinction law of 
\citet{car89} are shown for $R_V=3.1$ and 5 ({\it dashed lines}).
A higher value of $R_V$ is implied by the data for Chamaeleon~I,
which can be fit by a slope of 2.5 ({\it solid line}).
}
\label{fig:av}
\end{figure}

I now compare the color excesses $E(J-H)$ and $E(0.6-0.9)$ by plotting
them against each other in Figure~\ref{fig:av}. 
The fact that the distribution of these data intersects the origin indicates 
that the intrinsic $J-H$ colors and optical spectra that have been assumed are
reasonably accurate, on average. In other words, the members with the 
bluest spectra from the previous paragraph indeed have negligible extinction
according to their $J-H$ colors. 
In Figure~\ref{fig:av}, I next show the relations between $E(J-H)$ and 
$E(0.6-0.9)$ that are produced by the extinction law from \citet{car89} for
$R_V\equiv A_V/E(B-V)=3.1$ and 5. $R_V=3.1$ is the standard ratio observed
in the diffuse interstellar medium while $R_V=5$ is a relatively high value 
found in some dense clouds. For instance, several previous studies have 
reported ratios as high as $R_V=5$-6 for stars toward Chamaeleon~I
\citep{gra75,ryd80,vr84,ste89,whi94,whi87,whi97}. Indeed, 
the data in Figure~\ref{fig:av} qualitatively confirm the 
presence of high values of $R_V$, although there is no reasonable value of
this ratio for which the extinction law from \citet{car89} reproduces the
observed relation of $E(0.6-0.9)/E(J-H)=2.5$ derived here for Chamaeleon~I.
Because the effective wavelengths of photometric bands shift to longer
wavelengths as extinction increases, the constant in this relation is 
expected to change with extinction. However, upon simulating this effect
with the system transmissions for 2MASS $J$ and $H$, I find that this slope 
should change by only a few percent across the range of extinctions 
considered in Figure~\ref{fig:av}. Thus, the fitting of a linear relation 
between $E(0.6-0.9)$ and $E(J-H)$ is valid here.

Finally, the measurements of $E(J-H)$ and $E(0.6-0.9)$ are converted to 
extinctions.
Because \citet{car89} found that the ratio $E(J-H)/A_J$ did not change with 
different values of $R_V$, this relation should be applicable to Chamaeleon~I
regardless of anomalies in its extinction law. 
However, $E(J-H)/A_J$ does differ between the extinction laws of \citet{car89} 
and \citet{rl85}, who derive 0.35 and 0.38 for this ratio, respectively. 
In this work, I adopt $A_J=E(J-H)/0.38$, which I combine with 
$E(0.6-0.9)/E(J-H)=2.5$ to arrive at $A_J=E(0.6-0.9)/0.95$.
For each member of Chamaeleon~I, the final estimate of $A_J$ is the average
of the values produced by $E(J-H)$ and $E(0.6-0.9)$ when both are available, 
and otherwise is the value from the one measured color excess.

\subsection{Effective Temperatures and Bolometric Luminosities}
\label{sec:teff}

Spectral types of M0 and earlier are converted to effective temperatures 
with the dwarf temperature scale of \citet{sk82}.
For spectral types later than M0, I use the temperature scale that was designed 
by \citet{luh03b} to be compatible with the models of \citet{bar98} and 
\citet{cha00}. Bolometric luminosities are estimated by combining dereddened 
$J$-band measurements, a distance of 168~pc \citep{whi97,wic98,ber99}, and 
bolometric corrections described in \citet{luh99}.

The extinctions, effective temperatures, bolometric luminosities, and adopted 
spectral types for the members of Chamaeleon~I are listed in
Table~\ref{tab:mem2}.

\subsection{H-R Diagram}
\label{sec:hr}

The effective temperatures and bolometric luminosities for the members of
Chamaeleon~I can be interpreted in terms of masses and ages with theoretical 
evolutionary models. After considering the available sets of models, 
\citet{luh03b} concluded that those of \citet{pal99} for $M/M_\odot>1$ and 
\citet{bar98} and \citet{cha00} for $M/M_\odot\leq1$
provided the best agreement with observational constraints.
The spectroscopically classified members of Chamaeleon~I are plotted with 
these models on the H-R diagram in Figure~\ref{fig:hr}.
The cluster sequence is bounded by the isochrones for 1 and 10~Myr 
across the entire mass range in the current census.
It is unclear whether a spread of ages is actually reflected in these data
as the finite width of the sequence could also reflect extinction uncertainties,
unresolved binaries, and variability from accretion and rotation of 
spotted surfaces \citep{kh90,har01}.
As long as unresolved binaries are not the dominant component of the
width of the cluster sequence, the median of the sequence should reflect
the median age of the population, which is a useful characteristic age
that can be compared among young populations. In Figure~\ref{fig:hr}, the
models indicate a median age of $\sim2$~Myr for Chamaeleon~I, which is 
similar to that derived for IC~348 with these models \citep{luh03b} and 
slightly greater than the age of $\sim1$~Myr for Taurus \citep{bri02,luh03a}.
Chamaeleon~I and IC~348 also exhibit comparable 
fractions of sources with IR excess emission \citep{kg01,hai01}, which is 
consistent with a common evolutionary state.

T14A, CHSM~15991, and ISO~225 are anomalously faint for their spectral types, 
falling below the main sequence in Figure~\ref{fig:hr}.
Field stars that are mistakenly assigned membership in a young cluster can 
exhibit this characteristic, but youth and membership were firmly 
established for these stars with the diagnostics in \S~\ref{sec:class}. 
Alternatively, if these sources are occulted by circumstellar structures 
(e.g., edge-on disks), their observed photometry would measure only scattered 
light and produce an underestimate of the luminosity. 
In the case of the late-type object LS-RCrA~1 (M6.5),
\citet{fer01} suggested an additional explanation for its low luminosity 
in which intense accretion causes the evolutionary path of a low-mass star
or brown dwarf on the H-R diagram to depart from that predicted by models 
of non-accreting objects. \citet{com03} favored this scenario more strongly 
in their discussion of another subluminous late-type object, Par-Lup3-4 (M5).
They argued that the H$\alpha$, He~I, and Ca~II emission lines in this source
and in LS-RCrA~1 are produced by accretion and therefore originate near the 
stellar surface rather than in an extended region, which would in turn imply 
that the high ratios of lines to continua are intrinsic to the systems rather 
than resulting from preferential occulting of the central stars versus the line 
emitting regions. 
\citet{com03} did point out that the presence of intrinsically intense lines
in Par-Lup3-4 and LS-RCrA~1 would not rule out occultations as the
cause of the low luminosity estimates, but they deemed this possibility 
unlikely given the apparent absence of non-occulted counterparts, namely 
late-type objects with comparably strong emission lines and normal luminosities.
I find the following problems with this line of reasoning.
First, in LS-RCrA~1 and Par-Lup3-4, \citet{fer01} and \citet{com03} 
derived 6.7 and 11, respectively, for the ratio of H$\alpha$ to the sum 
of the [S~II] lines at 6714 and 6729~\AA. Those authors used the high 
values of this ratio as evidence that the H$\alpha$ emission is
produced predominantly by accretion, and suggested that comparable ratios 
are not observed for stars occulted by edge-on disks. 
However, the spectrum of the edge-on system MBM~12A~3C 
exhibits an even higher ratio of $\sim13$ \citep{jay02}. Thus, even when a star 
is completely occulted, the observed H$\alpha$ emission can arise primarily
from the accretion zone. This result is supported further by data on the young 
star KH15D, which is eclipsed periodically by circumstellar material
\citep{ham01}. Over the course of an eclipse, 
\citet{ham03} found that the stellar continuum changed by a factor of $\sim25$
while the line flux in H$\alpha$, whose line profile was indicative of
accretion, remained constant within a factor of two. These data demonstrate
that a stellar photosphere can be occulted to a much greater degree than the
H$\alpha$-emitting portion of its accretion columns, which implies that the 
emission lines may not be intrinsically intense in Par-Lup3-4 and LS-RCrA~1.
Indeed, high-resolution spectroscopy of the emission lines in LS-RCrA~1
indicates that its accretion is not unusually strong and instead is comparable 
to that of the other young members of its cluster \citep{bar03}.
Next, contrary to the statements by \citet{com03}, young late-type objects 
have been observed with both normal luminosities and $L_{H\alpha}/L_{\rm bol}$
comparable to that in LS-RCrA~1 and Par-Lup3-4, examples of which are ISO~252 
in Chamaeleon~I (this work), source 621 in IC~348 \citep{luh03b}, and 
S~Ori~71 in $\sigma$~Orionis \citep{bar02}.
In addition, \citet{fer01} and \citet{com03} suggested that primarily low-mass
sources would have their evolutionary paths altered by intense accretion, 
whereas subluminous sources have been found across a wide range of masses 
in Chamaeleon~I, IC~348, and Taurus 
\citep{luh03b,bri02}. For instance, like LS-RCrA~1 and Par-Lup3-4, T14A 
exhibits anomalously faint photometry and intense emission lines that are 
indicative of accretion (H$\alpha$/[S~II]$\sim8$, Ca~II triplet $\sim1$:1:1),
but its spectral type implies a much higher mass ($\sim0.8$~$M_{\odot}$).
In summary, while it is possible that the low luminosities of some young 
low-mass objects are due to intense accretion as proposed by \citet{fer01} and 
\citet{com03}, all available data for sources of this kind are consistent with 
a known phenomenon -- the occultation by circumstellar material. Consequently,
there is currently no motivation for invoking a second mechanism that may or 
may not occur in nature.

Recent studies (e.g., \citet{gp02,gm03}) have attempted to
validate stellar properties inferred from H-R diagrams 
through comparison to the results of \citet{law96}.
However, some of the stars assumed to be members by \citet{law96} 
have been shown to be field stars in this work.
Those authors also estimated spectral types for many of their sources from 
broad-band photometry alone (which unfortunately have been adopted as 
bonafide spectral types by some subsequent authors). 
In addition, \citet{law96} did not provide a quantitative analysis of 
completeness during the derivation of the IMF and their adopted models are 
now outdated. For these reasons, the masses, ages, IMF, and star formation 
history reported by \citet{law96} have questionable accuracy. The study by 
\citet{gm03} suffers from similar problems, in contrast to the careful 
analysis of the IMF in Chamaeleon~I by \citet{com99} and \citet{com00}.

\section{Conclusions}

As a part of a new census of the Chamaeleon~I star-forming region, 
I have performed optical spectroscopy on most of the sources that 
have been previously identified as possible members of the cluster, that 
lack either accurate spectral types or clear evidence of membership, 
and that are optically visible ($I\lesssim18$).
After measuring spectral types for the 179 objects in this sample, 
I used the spectroscopic data and all other available constraints to 
evaluate the membership status of 288 potential members that have been 
presented in published surveys.
This analysis has produced a list of 158 confirmed members, 8 of which are 
later than M6 and thus are likely to be brown dwarfs according to the 
evolutionary models of \citet{bar98} and \citet{cha00}. The membership of
41 of these sources is established for the first time by data in this paper.
Many of the objects that have been referred to as members of Chamaeleon~I 
in previous work lack evidence of membership. 
For instance, I find that approximately half of the candidates that were 
classified as young stars through IR photometry and spectroscopy by 
\citet{per00} and \citet{gm03}, respectively, are field stars, predominantly 
background giants. 
Meanwhile, most of the candidates discovered by \citet{car02} are
spectroscopically confirmed as members, one of which is the coolest known 
member of Chamaeleon~I at a spectral type of M8 ($\sim0.03$~$M_{\odot}$).

For the known members of Chamaeleon~I that have accurate spectral types,
I have estimated extinctions, luminosities, and effective temperatures and
used these data to construct an H-R diagram for the cluster. 
Evolutionary models imply a median age of $\sim2$~Myr for Chamaeleon~I,
which is similar to that of IC~348 and slightly greater than the age of 
$\sim1$~Myr for Taurus.
In addition to ages, masses of the members of Chamaeleon~I can be 
estimated from the H-R diagram and evolutionary models. 
However, the current census of the cluster is not suitable for deriving an IMF
because the known members were originally identified by signatures 
of youth (H$\alpha$, IR excess, variability) that have different and 
ill-defined sensitivities as a function of mass. Indeed,
I find that it is not possible to define an area of the cluster and an
extinction limit within which the current census contains a significant
number of members and approaches a high level of completeness for a useful 
range of masses, which is necessary for constructing a mass function that is 
unbiased in mass and thus a meaningful representation of the cluster. 
The measurement of an IMF for Chamaeleon~I from spectroscopic data will
require a magnitude-limited survey for cluster members.

\acknowledgements

I thank the staff at Las Campanas Observatory, particularly Mauricio Navarrete,
for their outstanding support of these observations and an anonymous referee 
for suggested improvements to the manuscript.
I am grateful to Isabelle Baraffe and Francesco Palla for access to their
most recent calculations and to Laurent Cambr\'esy for providing the extinction
map for Chamaeleon. I was supported by grant NAG5-11627 from the 
NASA Longterm Space Astrophysics program.
This research has made use of the NASA/IPAC Infrared Science Archive, which is
operated by the Jet Propulsion Laboratory, California Institute of Technology,
under contract with the National Aeronautics and Space Administration.

\appendix

\section{Comments on Individual Sources}
\label{sec:append}

\begin{figure}
\plotone{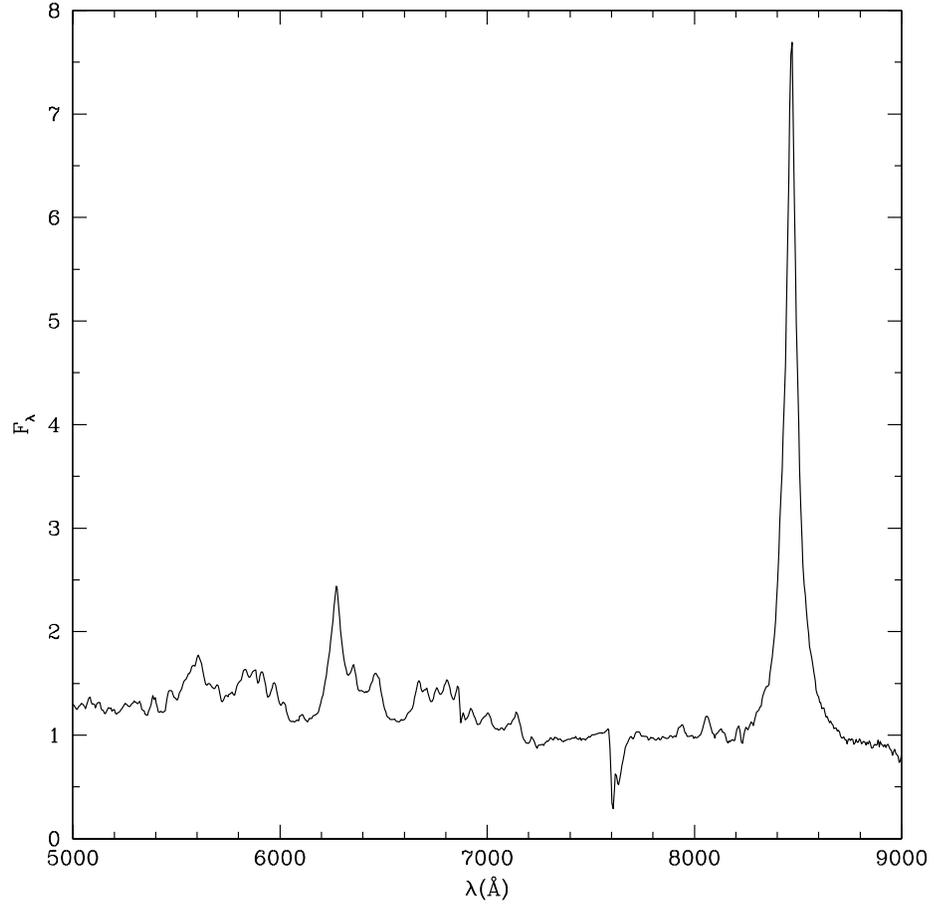}
\caption{Low-resolution spectrum of CHSM~11564, which exhibited $K$-band 
excess emission in the data of \citet{car02} and is located $1\fdg5$
south of Chamaeleon~I. 
The wavelengths and widths (FWHM$\sim3000$~km~s$^{-1}$) of the two strongest 
emission lines are consistent with H$\alpha$ and H$\beta$ arising from a 
Seyfert~1 galaxy at a redshift of 0.29.
The data are displayed at a resolution of 13~\AA\ and are normalized at
7500~\AA.}
\label{fig:fb5}
\end{figure}

\citet{car02} detected $K$-band excess emission in CHSM~11564, which is 
located more than a degree from the edge of the Chamaeleon~I clouds. 
They suggested that this object is either an isolated young star or a bright,
red galaxy. Indeed, the spectrum of CHSM~11564 in Figure~\ref{fig:fb5}
is indicative of a Seyfert~1 galaxy (Balmer FWHM$\sim3000$~km~s$^{-1}$) 
at a redshift of 0.29.

It appears that \citet{car02} mistakenly matched Cam2-26 to 
2MASS~11073775-7735308, or Cha~H$\alpha$~7, instead of the true counterpart, 
2MASS~11073677-7735167.
Also, the correct match for Ced112-IRS2 is C1-6 rather than C1-3.
\citet{per00} associated ISO~274 with the DENIS source that corresponds to
2MASS~11112550-7706100, but noted that this identification was uncertain 
because of the presence of multiple near-IR sources in the vicinity of the 
position of ISO~274.
After classifying 2MASS~11112550-7706100 as a field star through spectroscopy,
I observed one of the other near-IR sources close to the position of ISO~274, 
2MASS~11112260-7705538, which I classified as a member of Chamaeleon~I.
Therefore, ISO~274 is matched to 2MASS~11112260-7705538 in this work.

\begin{figure}
\epsscale{0.8}
\plotone{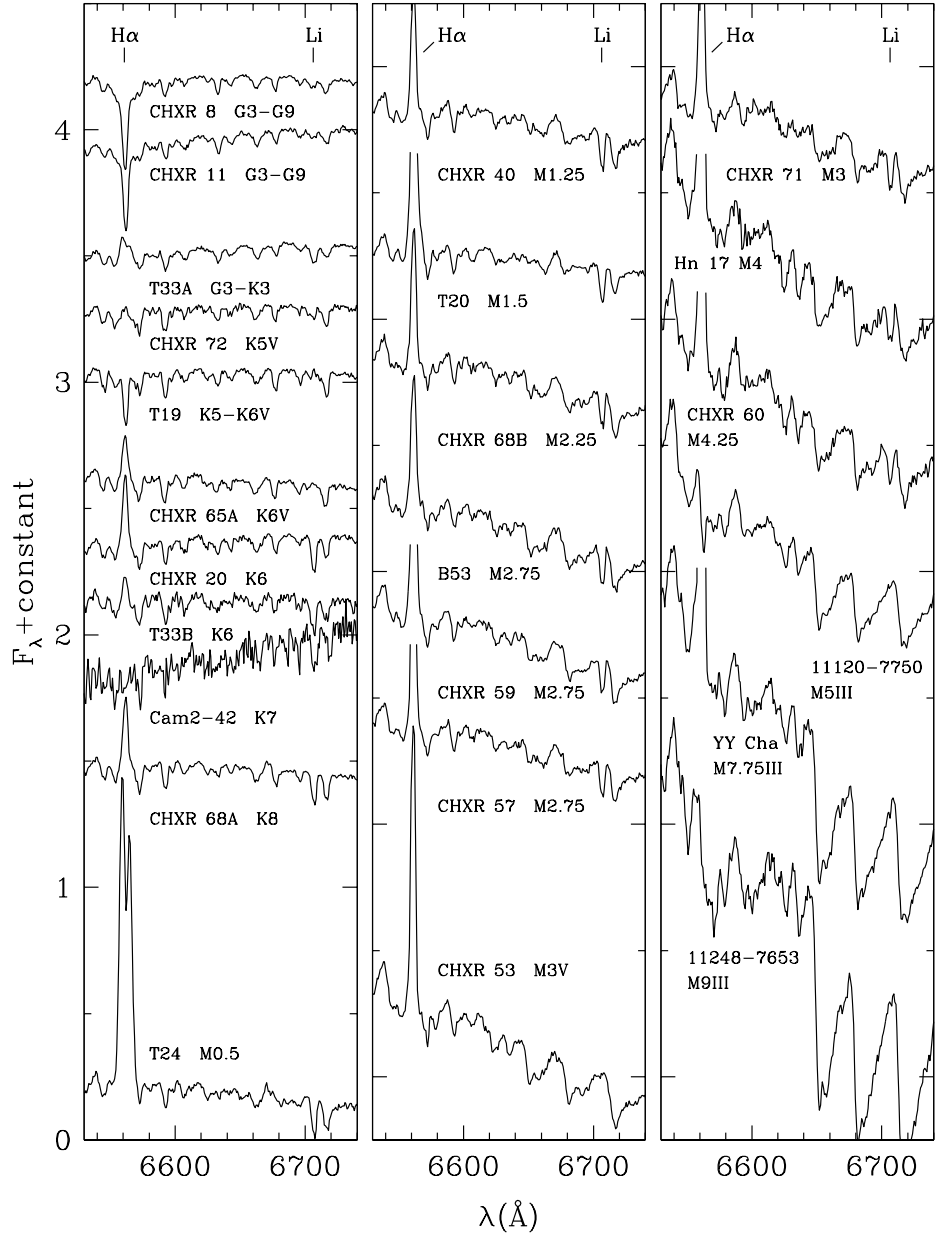}
\caption{Medium-resolution spectra of stars toward the Chamaeleon~I 
star-forming region. Strong absorption in Li 6707~\AA\ is evidence of 
youth and thus probable membership in Chamaeleon~I. Stars in which Li is 
weak or absent are field dwarfs and giants. 
The luminosity classes are uncertain for the probable field stars CHXR~8 and 11.
These data have a resolution of 3~\AA\ and are normalized to the
continuum near the Li line.}
\label{fig:li}
\end{figure}

\citet{fk89} resolved T33 into a visual double with a separation of 
$\sim4\arcsec$ and obtained optical spectroscopy and photometry for the
individual components. The later, brighter, and bluer star was designated 
as ``A" and the other star as ``B". From the spectra in this work, I
instead find that the earlier (eastern) star is brighter across the spectral 
range shown in Figure~\ref{fig:li}. This source also was found to be the
brighter component at $K$ by \citet{ghe97}. Therefore, I refer to it as T33A. 
This source probably corresponds to the B component from \citet{fk89}, but 
a conclusive cross-identification is not possible because the relative 
positions of the two stars were not described by \citet{fk89}.
The $J$, $H$, and $K_s$ data for T33A+B are from \citet{car02} since $H$ and
$K_s$ are not available from the 2MASS Point Source Catalog. 

The proper motion of HD~94216 is not consistent with membership in Chamaeleon~I
\citep{hog98}.
The weak Li absorption ($W_{\lambda}\sim0.1$~\AA) in the spectrum of CHXR~8 
is indicative of a field star (this work, \citet{hlf94}).
Because of the uncertainty in the luminosity class of each of these stars
and the lack of significant reddening in their colors, they could be either 
in the foreground or the background of Chamaeleon~I. HD~94216 and CHXR~8 are 
listed as background stars for the purposes of this work.

\citet{per00} identified HM8 (T12) as the counterpart to their mid-IR source
ISO~10. In the data that they listed for HM8, I find that the $iJK_s$ 
measurements apply to HM8 while the coordinates are instead for a nearby
fainter star, 2MASS~11024866-7721453. The difference between those coordinates
and the correct values for HM8 led \citet{car02} to question the matching
of ISO~10 to HM8 and thus list them as separate sources in their compilation.
However, because \citet{per00} reported the coordinates of the matched near-IR 
stars rather than the measurements from their {\it ISO} images, the matching
between {\it ISO} and near-IR sources cannot be verified in that way.
In this work, I assume that the unpublished {\it ISO} coordinates of 
ISO~10 agree with the near-IR coordinates of HM8, which resulted in the
matching of these sources by \citet{per00}, and that these authors 
provided the correct near-IR photometry but the wrong coordinates for HM8.

C1-14/F32/CHX~15A/ISO~205 cannot be in the foreground of Chamaeleon~I 
given that it is reddened \citep{whi87,fk89,whi97}. 
If it is a background star, the 
combination of a dwarf classification \citep{vr84} and a position near 
the main sequence for the distance of Chamaeleon~I \citep{whi87,fk89,whi97} 
indicates that it must be very close to the opposite site of the cloud.
C1-14 appears to exhibit excess emission in the $M$-band data of \citet{pru92}
but not in the 6.7 and 14.3~\micron\ photometry of \citet{per00}. 
It is unclear whether this star is a background dwarf or a member of 
Chamaeleon~I.

HD~93828 is on the outskirts of Chamaeleon~I at a projected 
distance of $\sim1\arcdeg$ from the cluster center. The parallax and proper 
motion of this star are consistent with membership \citep{per97}. 
In addition, HD~93828 is modestly redder in $H-K_s$ ($\sim0.1$~mag) than 
expected for a reddened F0V star, which is suggestive of IR excess emission and 
youth. For these reasons, this star is treated as a cluster member in this work.

\begin{figure}
\plotone{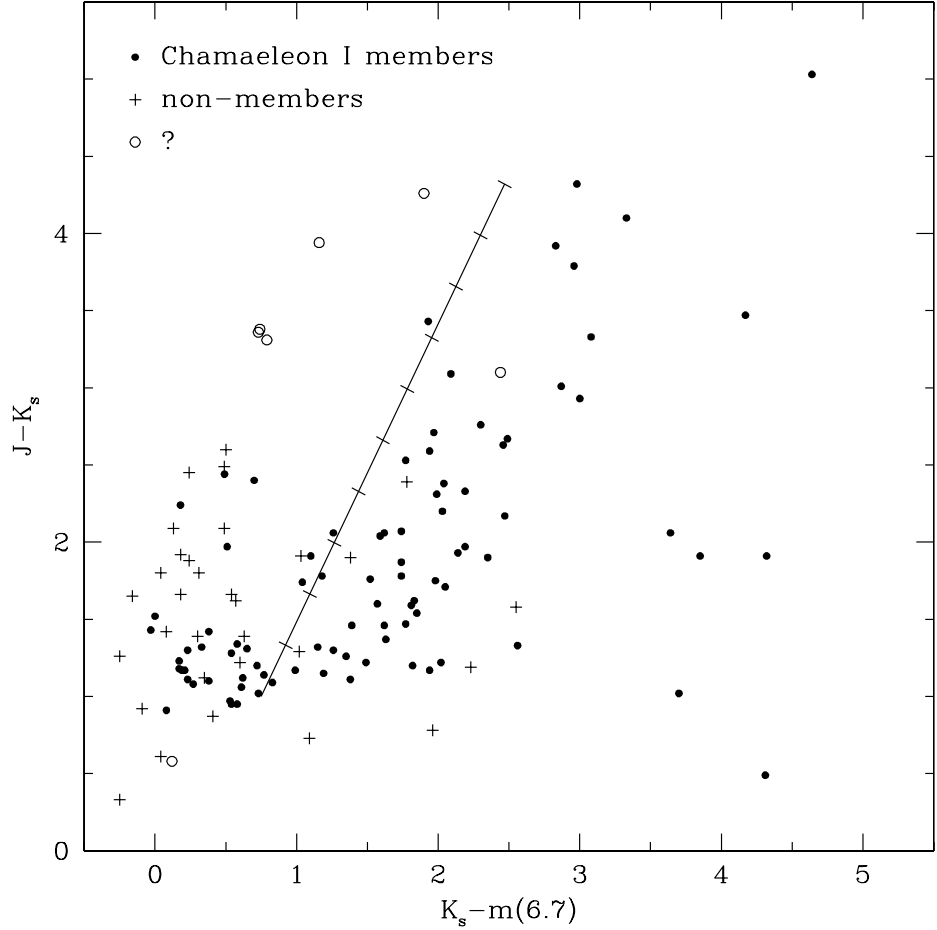}
\caption{
$K_s-m(6.7)$ versus $J-K_s$ for objects that are classified as members 
({\it points}) and as non-members ({\it plusses}) of the Chamaeleon~I 
star-forming region (\S~\ref{sec:confirm}). 
The solid line is a reddening vector placed at an arbitrary origin and marked
at intervals of $A_V=2$.
One of the sources whose membership could not be determined ({\it circles}; 
Table~\ref{tab:unc}) is redder in $K_s-m(6.7)$ than expected
for a typical reddened field star and therefore may exhibit IR excess emission
that would indicate youth and membership in Chamaeleon~I.
These measurements are from \citet{per00} ($m(6.7)$) and 2MASS ($J$, $K_s$).}
\label{fig:iso2}
\end{figure}

In \citet{per00}, it was unclear whether the correct near-IR counterpart had
been identified for ISO~68. 
\citet{com00} supported this uncertainty when they did not detect H$\alpha$
emission in the near-IR source assigned to ISO~68.
I obtained spectra for this star and for the two nearest near-IR 
sources as well, which I refer to as ISO~68A, B, and C. 
All of these objects are field stars according to my data. 
ISO~68 is either one of these field stars or a highly embedded source that 
is not detected by 2MASS. Since ISO~68 is detected only at 6.7~\micron\ and 
not at 14.3~\micron, there is no evidence that it is particularly red 
and thus no reason to adopt the latter explanation. 
Instead, if ISO~68A is the true near-IR counterpart, then its colors 
($J-K_s=1.62$, $K-6.7=0.57$) do not exhibit significant excess emission 
beyond that expected from a reddened field star (Figure~\ref{fig:iso2}).

During my spectroscopy of the near-IR source matched to ISO~250 by
\citet{per00}, 2MASS~11103481-7722053 happened to fall in the slit as well. 
Both of these stars are classified as members of Chamaeleon~I based
on the spectra. Given the separation of $9\arcsec$ between these two stars,
it is likely that \citet{per00} correctly identified the near-IR counterpart 
to the mid-IR source ISO~250.

Although HD~96675 is not commonly included in lists of members of Chamaeleon~I
in previous studies, it is a source of X-ray emission \citep{fei93}, 
which is suggestive of youth, and has a parallax and a proper motion that 
are consistent with membership in Chamaeleon~I \citep{ber99,ter99,tei00}. 
Therefore, HD~96675 is considered a cluster member in this work.

T1, T18, and Hn14 were originally detected as weak sources of 
H$\alpha$ emission in objective prism surveys \citep{sch77,har93}. However, no 
evidence of youth has been presented for these stars in subsequent studies.
Their spectra exhibit absorption rather than emission in H$\alpha$ and are 
indicative of background stars, probably giants. 

The field dwarfs CHXR~65A and 65B appear to have a common distance of 120~pc 
according to their spectral types and photometry, and thus are probably 
components of a binary system.

\begin{figure}
\plotone{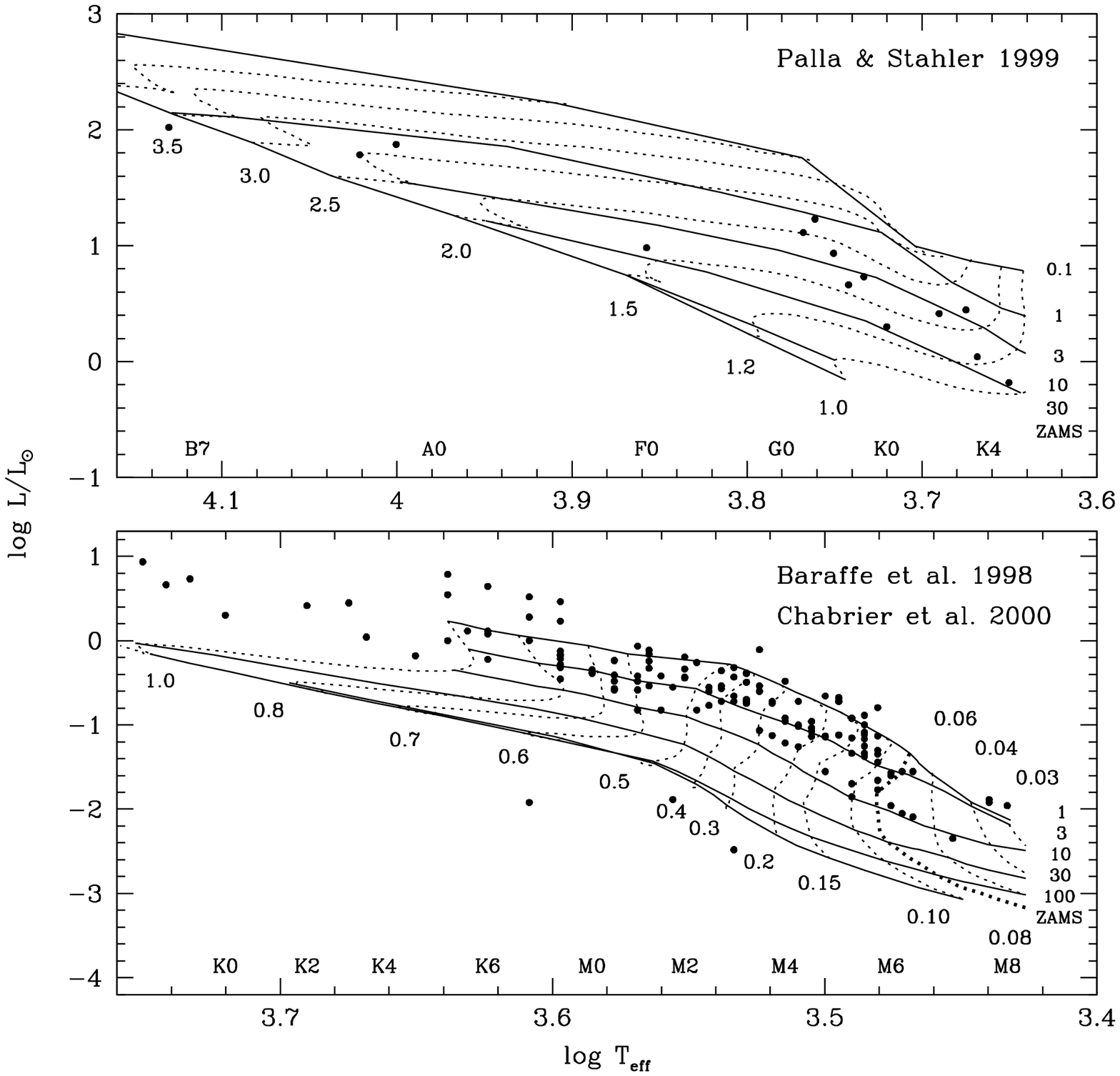}
\caption{
H-R diagram for the members of the Chamaeleon~I star-forming region that have
measured spectral types.
These data are shown with the theoretical evolutionary models of \citet{pal99} 
({\it upper panel}) and \citet{bar98} ($0.1<M/M_\odot\leq1$) and \citet{cha00} 
($M/M_\odot\leq0.1$) ({\it lower panel}), where the mass tracks 
({\it dotted lines}) and isochrones ({\it solid lines}) are labeled in units 
of $M_\odot$ and Myr, respectively.
The cluster members that are below the main sequence (T14A, ISO~225, CHSM~15991)
are probably detected primarily in scattered light, which precludes the 
measurement of accurate luminosities.
}
\label{fig:hr}
\end{figure}

Some previous studies have assumed that the B9 star T41/HD~97300 is on the 
zero-age main sequence. For instance, based on this assumption, \citet{whi97} 
derived a distance of 152~pc for the star, and hence for the cloud. However, 
a star with a spectral type of B9 is not necessarily old enough to have reached 
the main sequence in a cluster like Chamaeleon~I, as illustrated by the 
the evolutionary models of \citet{pal99} in Figure~\ref{fig:hr}. 
Indeed, T41 is above the main sequence in that H-R diagram and yet has an 
inferred age that is within the range exhibited by the other cluster members.

\begin{figure}
\plotone{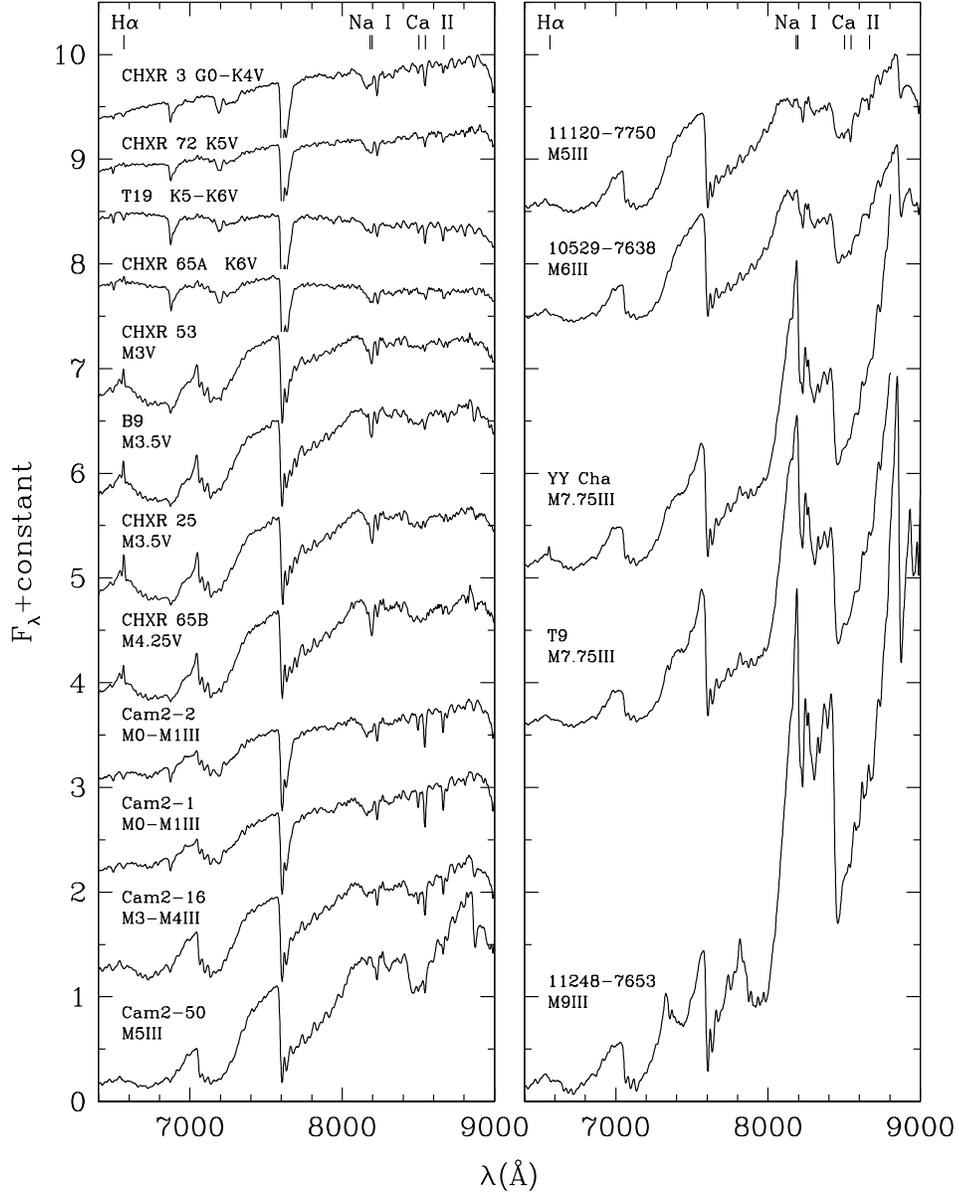}
\caption{Low-resolution spectra of objects toward the Chamaeleon~I star-forming
region that are classified as late-type field dwarfs and giants
(\S~\ref{sec:confirm}). 
The data are displayed at a resolution of 13~\AA\ and are normalized at
7500~\AA.}
\label{fig:fb4}
\end{figure}

Although YY~Cha exhibits excess emission at mid-IR wavelengths, the low
level of this excess led \citet{whi91} to support its original classification 
as a Mira field star from \citet{kho85}. This classification is confirmed
through the spectrum of YY~Cha in Figure~\ref{fig:fb4}. 
Similarly, T9, IRAS~11248-7653, IRAS~11120-7750, and IRAS~10529-7638 were
identified as possible members of Chamaeleon~I based on their mid-IR 
excesses \citep{whi91,gs92,pru92}, but are found to be late-M giants with the 
spectra in this work.
While H$\alpha$ emission is not observed in most field giants, it can appear 
in Mira variables \citep{joy26} and its presence in the data for YY~Cha
is not an indication of youth.

\citet{dub96} found that the radial velocity of T19 was similar to the 
velocities of known members of Chamaeleon~I and the molecular gas in the cloud,
which implied that it was a member of the cluster. However, the absence of 
Li absorption ($W_{\lambda}<0.1$~\AA) in the spectrum of T19 in 
Figure~\ref{fig:li} indicates that it is a field dwarf. It probably is in the 
foreground of Chamaeleon~I given the lack of significant reddening in its
colors and spectrum ($A_V<1$). The low rotational velocity of T19 from 
\citet{dub96} (V~sin~i$<7.9$~km~s$^{-1}$) is consistent with those of old field 
dwarfs. CHXR~25/B33 is another star whose radial velocity is suggestive of 
membership \citep{cov97} but is classified as a field dwarf by the diagnostics 
in \S~\ref{sec:class}.

During the LDSS-2 observations, a direct image without a filter was obtained 
to identify each target for spectroscopy. Through this imaging, CHXR~26 
was resolved into a $1\farcs4$ double in which the components
have similar brightnesses.

\clearpage

\begin{deluxetable}{llllll}
\setlength{\tabcolsep}{0.04in}
\tabletypesize{\scriptsize}
\rotate
\tablewidth{0pt}
\tablecaption{Observing Log \label{tab:log}}
\tablehead{
\colhead{Date} &
\colhead{Telescope + Instrument} &
\colhead{Grating} &
\colhead{Resolution} &
\colhead{ID} \\
\colhead{} &
\colhead{} &
\colhead{l~mm$^{-1}$} &
\colhead{\AA} &
\colhead{}
}
\startdata
2003 Mar 20 & Magellan~II + LDSS-2 & 300 & 13 & CHSM~1715, CHSM~1982, CHSM~7869, CHSM~9484, CHSM~10193, CHSM~10862, \\
\nodata & \nodata & \nodata & \nodata & CHSM~11564, CHSM~15991, CHSM~17173, GL~406, LHS~2065, VB~8 \\
2003 Mar 21 & Magellan~II + LDSS-2 & 300 & 13 & Cha~H$\alpha$~1, Cha~H$\alpha$~7, Cha~H$\alpha$~10, Cha~H$\alpha$~11, Cha~H$\alpha$~12, CHXR~65B, ISO~28, \\
\nodata & \nodata & \nodata & \nodata & ISO~114, ISO~138, ISO~145, ISO~147, ISO~220, ISO~252, LHS~2065 \\
2003 Mar 22 & Magellan~II + LDSS-2 & 300 & 13 & Cha~H$\alpha$~2, Cha~H$\alpha$~3, Cha~H$\alpha$~4, Cha~H$\alpha$~5, Cha~H$\alpha$~6, Cha~H$\alpha$~8, Cha~H$\alpha$~9, \\
\nodata & \nodata & \nodata & \nodata & Cha~H$\alpha$~13, CHXR~22E, CHXR~22W, CHXR~26, Hn12E, Hn14, Hn21E, Hn21W, \\
\nodata & \nodata & \nodata & \nodata & ISO~68A, ISO~68B, ISO~68C, ISO~71, ISO~79, ISO~143, ISO~154, ISO~164, ISO~165, \\
\nodata & \nodata & \nodata & \nodata & ISO~217, ISO~225, ISO~235, ISO~247, ISO~274, ISO~282, OTS~42, VB~8 \\
\enddata
\end{deluxetable}

\begin{deluxetable}{lllll}
\tablewidth{0pt}
\tablecaption{H$\alpha$ for Members of Chamaeleon~I \label{tab:ha}}
\tablehead{
\colhead{ID} &
\colhead{$W_{\lambda}$} &
\colhead{Flux\tablenotemark{a}} &
\colhead{log~$L_{H\alpha}/L_{\rm bol}$\tablenotemark{b}} &
\colhead{Spectral Type\tablenotemark{c}} \\
\colhead{} &
\colhead{(\AA)} &
\colhead{} &
\colhead{} &
\colhead{}}
\startdata
      T4 & 7.5$\pm$0.5 &        210 &      -3.36 &       M0.5  \\
      T5 &   22$\pm$1 &        150 &      -3.05 &      M3.25  \\
      T7 &   41$\pm$2 &        900 &      -2.68 &         K8  \\
     T10 &   67$\pm$2 &        130 &      -2.74 &      M3.75  \\
  CHXR 9C & 3.4$\pm$0.3 &         60 &      -3.71 &      M2.25  \\
     T11 &   20$\pm$1 &       2200 &      -2.77 &         K6  \\
  CHXR 71 &  4$\pm$0.3 &         13 &      -3.78 &         M3  \\
     T12 &   45$\pm$1 &        100 &      -2.98 &       M4.5  \\
\enddata
\tablenotetext{a}{Observed line flux in units of 
$10^{-15}$~erg~s$^{-1}$~cm$^{-2}$. Because the $i$-band photometry used 
for flux calibration and the spectroscopic data were not obtained at the same
time, the dominant source of error for most of these measurements is 
variability, which is typically $\pm10$\% for young stars \citep{car01}.}
\tablenotetext{b}{Typical uncertainties are $\pm0.1$.}
\tablenotetext{c}{From this work.}
\end{deluxetable}

\begin{deluxetable}{lll}
\tablewidth{0pt}
\tablecaption{Li 6707 for Members of Chamaeleon~I \label{tab:li}}
\tablehead{
\colhead{ID} &
\colhead{$W_{\lambda}$} &
\colhead{Spectral Type\tablenotemark{a}} \\
\colhead{} &
\colhead{(\AA)} &
\colhead{} 
}
\startdata
  CHXR 71 & 0.55$\pm$0.05 &         M3  \\
     T20 & 0.57$\pm$0.05 &       M1.5  \\
  CHXR 20 & 0.65$\pm$0.1 &         K6  \\
     T24 & 0.59$\pm$0.05 &       M0.5  \\
    T33A & 0.32$\pm$0.05 &         G3-K3  \\
    T33B & 0.60$\pm$0.05 &         K6  \\
 Cam2-42 & 0.77$\pm$0.1 &         K7  \\
  CHXR 40 & 0.62$\pm$0.05 &      M1.25  \\
    Hn17 & 0.65$\pm$0.05 &         M4  \\
  CHXR 57 & 0.53$\pm$0.05 &      M2.75  \\
  CHXR 59 & 0.48$\pm$0.05 &      M2.75  \\
  CHXR 60 & 0.60$\pm$0.05 &      M4.25  \\
     B53 & 0.70$\pm$0.05 &      M2.75  \\
 CHXR 68B & 0.53$\pm$0.05 &      M2.25  \\
 CHXR 68A & 0.46$\pm$0.05 &         K8  \\
\enddata
\tablenotetext{a}{From this work.}
\end{deluxetable}

\begin{deluxetable}{lllllllll}
\tabletypesize{\scriptsize}
\tablewidth{0pt}
\tablecaption{Astrometry and Photometry for Members of Chamaeleon I \label{tab:mem1}}
\tablehead{
\colhead{ID} &
\colhead{$\alpha$(J2000)\tablenotemark{a}} &
\colhead{$\delta$(J2000)\tablenotemark{a}} &
\colhead{$i$\tablenotemark{b}} & 
\colhead{$J-H$\tablenotemark{c}} & \colhead{$H-K_s$\tablenotemark{c}} 
& \colhead{$K_s$\tablenotemark{c}} &
\colhead{m(6.7)\tablenotemark{d}} & \colhead{m(14.3)\tablenotemark{d}}} 
\startdata
 HD 93828$\tablenotemark{e}$ &   10 46 37.95 &   -77 36 03.6 &     8.88 &     0.19 &     0.18 &     7.48 &       \nodata &       \nodata  \\
      T3$\tablenotemark{f}$ &   10 55 59.73 &   -77 24 39.9 &    12.60 &     0.81 &     1.15 &     8.69 &       \nodata &       \nodata  \\
      T4 &   10 56 30.45 &   -77 11 39.3 &    11.29 &     0.87 &     0.47 &     8.63 &       \nodata &       \nodata  \\
      T5 &   10 57 42.20 &   -76 59 35.7 &    12.20 &     0.87 &     0.32 &     9.25 &       \nodata &       \nodata  \\
      T6 &   10 58 16.77 &   -77 17 17.1 &    10.62 &     0.85 &     0.65 &     7.76 &       \nodata &       \nodata  \\
      T7 &   10 59 01.09 &   -77 22 40.7 &    11.68 &     0.90 &     0.62 &     8.62 &       \nodata &       \nodata  \\
      T8 &   10 59 06.99 &   -77 01 40.4 &     9.68 &     0.65 &     0.51 &     7.31 &       \nodata &       \nodata  \\
     T10 &   11 00 40.22 &   -76 19 28.1 &    13.64 &     0.62 &     0.37 &    10.87 &       \nodata &       \nodata  \\
\enddata
\tablenotetext{a}{2MASS Point Source Catalog, except for ISO~13, which is 
from \cite{car02}.}
\tablenotetext{b}{Second DENIS Release.}
\tablenotetext{c}{2MASS Point Source Catalog, except for $J$ for T3, which is
from the Second DENIS Release.}
\tablenotetext{d}{\citet{per00}.}
\tablenotetext{e}{Additional comments on this source are 
in \S~\ref{sec:append}.}
\tablenotetext{f}{
Sources T3, T27, T31, T33, T39, T41, T45, T51, T54, CHXR 47, and Cha~H$\alpha$~2
are resolved as binaries with separations of 
$2\farcs2$, $0\farcs8$, $0\farcs66$, $2\farcs4$-$2\farcs5$, $1\farcs1$,
$0\farcs84$, $0\farcs79$, $1\farcs5$-$1\farcs9$, $0\farcs27$, $1\farcs2$, and
$0\farcs2$ \citep{rz93,ghe97,neu02}.}
\end{deluxetable}

\begin{deluxetable}{lllllllll}
\tabletypesize{\scriptsize}
\rotate
\tablewidth{0pt}
\tablecaption{Derived Properties for Members of Chamaeleon I \label{tab:mem2}}
\tablehead{
\colhead{ID} &
\colhead{Spectral Type/$W_{\lambda}$(H$\alpha$)\tablenotemark{a}} &
\colhead{Ref} &
\colhead{Adopt} &
\colhead{Membership\tablenotemark{b}} &
\colhead{Ref} &
\colhead{$T_{\rm eff}$\tablenotemark{c}} &
\colhead{$A_J$} & \colhead{$L_{\rm bol}$} \\
\colhead{} &
\colhead{} &
\colhead{} &
\colhead{} &
\colhead{Evidence} &
\colhead{} &
\colhead{} &
\colhead{} &
\colhead{} 
}
\startdata
 HD 93828 &                  F0V &   1 &         F0 &    $\mu$,$\pi$ &    2 &   7200 &    0.00 &     9.6  \\
      T3 & M1,M0.5/14,M0.5/26.7 & 3,4,5 &       M0.5 &   e,ex,rv & (6,7,3,4,5,8,9,10),(11,12,13),14 &   3778 &    0.00 &    0.27  \\
      T4 &     K2,M0:/63.6,M0.5 & 15,5,16 &       M0.5 & e,ex,Li,rv & (6,7,15,3,4,5,9,16),(11,12,13),17,14 &   3778 &    0.17 &    0.58  \\
      T5 &                M3.25 &    16 &      M3.25 & e,ex,rv,NaK & (7,9,16),12,18,16 &   3379 &    0.50 &    0.41  \\
      T6 &                 K0/7 &    4 &         K0 &      e,ex & (4,19,9),(11,12,13,20) &   5250 &    0.42 &     2.0  \\
      T7 &           K0:,M0:,K8 & 3,5,16 &         K8 & e,ex,Li,rv,$A_V$ & (6,7,15,3,4,5,9,16),(11,12,13),17,14,16 &   3955 &    0.17 &    0.52  \\
      T8 &  G8,K0,K2/42,K2/43.6 & 6,15,4,5 &         K2 &   e,ex,rv & (6,7,15,4,5,9,10),(11,12,13,20),18 &   4900 &    0.00 &     2.6  \\
     T10 &                M3.75 &    16 &      M3.75 &     e,NaK & (6,7,9,16),16 &   3306 &    0.10 &   0.075  \\
\enddata
\tablenotetext{a}{
Both spectral types and $W_{\lambda}$(H$\alpha$) are provided when taken
from previous studies. The measurements of H$\alpha$ from this work are 
listed separately in Table~\ref{tab:ha}.
The precision of the optical spectral types from
this work is $\pm0.5$ and 0.25 subclass for K and M types, respectively, 
unless noted otherwise.}
\tablenotetext{b}{Membership in Chamaeleon~I is indicated by $A_V\gtrsim1$ and
a position above the main sequence for the distance of Chamaeleon~~I (``$A_V$"),
strong emission lines (``e"),
Na~I and K~I strengths intermediate between those of dwarfs and giants (``NaK"),
strong Li absorption (``Li"),
IR excess emission (``ex"), 
or a proper motion (``$\mu$"), parallax (``$\pi$"), or radial velocity 
(``rv") that is similar to that of the known members of Chamaeleon~I.} 
\tablenotetext{c}{Temperature scale from \citet{sk82} ($\leq$M0) and
\citet{luh03b} ($>$M0).}
\tablerefs{
(1) \citet{hc75};
(2) \citet{per97};
(3) \citet{app79};
(4) \citet{ryd80};
(5) \citet{app83};
(6) \citet{hm73};
(7) \citet{sch77};
(8) \citet{whi87};
(9) \citet{har93};
(10) \citet{nat00};
(11) \citet{bau84};
(12) \citet{pru92};
(13) \citet{gs92};
(14) \citet{jg01};
(15) \citet{app77};
(16) this work;
(17) \citet{gre92};
(18) \citet{dub96};
(19) \citet{whi97};
(20) \citet{hen93};
(21) \citet{hlf94};
(22) \citet{per00};
(23) \citet{vr84};
(24) \citet{fk89};
(25) \citet{gur99};
(26) \citet{pru91};
(27) \citet{leh01};
(28) \citet{wal92};
(29) \citet{gm03};
(30) \citet{com99};
(31) \citet{com00};
(32) \citet{gp02};
(33) \citet{law96};
(34) \citet{nc99};
(35) \citet{whi91};
(36) \citet{ass90};
(37) \citet{bz97};
(38) \citet{saf03};
(39) \citet{alc95};
(40) \citet{cov97}.}
\end{deluxetable}

\begin{deluxetable}{lllllllllll}
\setlength{\tabcolsep}{0.04in}
\tabletypesize{\scriptsize}
\tablewidth{0pt}
\tablecaption{Foreground Stars \label{tab:fore}}
\tablehead{
\colhead{ID} &
\colhead{$\alpha$(J2000)\tablenotemark{a}} &
\colhead{$\delta$(J2000)\tablenotemark{a}} &
\colhead{Spectral Type} &
\colhead{Ref} &
\colhead{Foreground\tablenotemark{b}} &
\colhead{Ref} &
\colhead{$i$\tablenotemark{c}} & 
\colhead{$J-H$\tablenotemark{a}} & \colhead{$H-K_s$\tablenotemark{a}} 
& \colhead{$K_s$\tablenotemark{a}} \\
\colhead{} &
\colhead{} &
\colhead{} &
\colhead{} &
\colhead{} &
\colhead{Evidence} &
\colhead{} &
\colhead{} &
\colhead{} &
\colhead{} &
\colhead{}
}
\startdata
 HD 95916 &   11 01 20.04 &   -77 36 12.6 &                  F8V &   1,2 &    sp &      2 &     8.83 &     0.29 &     0.03 &     7.87  \\
RXJ 1101.5-7655 &   11 01 27.13 &   -76 55 30.7 &                  M4V &        3 &      Li &      3 &    13.69 &     0.62 &     0.30 &    11.34  \\
      B9 &   11 02 15.52 &   -77 10 59.4 &                M3.5V &        4 &     NaK &      4 &    15.01 &     0.59 &     0.31 &    12.65  \\
     F20 &   11 03 44.50 &   -77 46 11.1 &                  K3V &       5 &    sp & 5,2 &       \nodata &     0.51 &     0.11 &     8.93  \\
      T19$\tablenotemark{d}$ &   11 05 41.53 &   -77 54 44.1 &               K5-K6V &        4 &      Li &      4 &    12.36 &     0.62 &     0.15 &    10.61  \\
   CHX 6A &   11 05 50.38 &   -77 07 22.0 &                K7,M0 &   6,7 &      Li &      7 &    11.47 &     0.63 &     0.17 &     9.71  \\
   CHXR 25$\tablenotemark{d}$ &   11 07 33.02 &   -77 28 27.7 &      M3-M4/7.0,M3.5V &    8,4 &  Li,NaK &  9,4 &    13.16 &     0.61 &     0.25 &    10.76  \\
 HD 97240 &   11 09 18.16 &   -77 47 39.9 &                  F5V &   1,2 & sp,$\mu$,$\pi$ & 2,10,10 &     8.74 &     0.21 &     0.12 &     7.39  \\
\enddata
\tablenotetext{a}{2MASS Point Source Catalog.}
\tablenotetext{b}{Status as a foreground field star is indicated by 
$A_V<$ and either a dwarf spectral classification (``sp"), weak Li 
absorption (``Li"), strong Na~I and K~I absorption (``NaK"), or a proper 
motion (``$\mu$") or parallax (``$\pi$") that differs from that of the known 
members of Chamaeleon~I.}
\tablenotetext{c}{Second DENIS Release.}
\tablenotetext{d}{Additional comments on this source are 
in \S~\ref{sec:append}.}
\tablerefs{
(1) \citet{hc75};
(2) \citet{whi87};
(3) \citet{alc95};
(4) this work;
(5) \citet{vr84};
(6) \citet{fk89};
(7) \citet{wal92};
(8) \citet{cov97};
(9) \citet{law96};
(10) \citet{per97};
(11) \citet{whi97};
(12) \citet{hlf94}.}
\end{deluxetable}

\begin{deluxetable}{lllllllllll}
\setlength{\tabcolsep}{0.04in}
\tabletypesize{\scriptsize}
\tablewidth{0pt}
\tablecaption{Background Stars \label{tab:back}}
\tablehead{
\colhead{ID} &
\colhead{$\alpha$(J2000)\tablenotemark{a}} &
\colhead{$\delta$(J2000)\tablenotemark{a}} &
\colhead{Spectral Type} &
\colhead{Ref} &
\colhead{Background\tablenotemark{b}} &
\colhead{Ref} &
\colhead{$i$\tablenotemark{c}} & 
\colhead{$J-H$\tablenotemark{a}} & \colhead{$H-K_s$\tablenotemark{a}} 
& \colhead{$K_s$\tablenotemark{a}} \\
\colhead{} &
\colhead{} &
\colhead{} &
\colhead{} &
\colhead{} &
\colhead{Evidence} &
\colhead{} &
\colhead{} &
\colhead{} &
\colhead{} &
\colhead{}
}
\startdata
 HD 93601 &   10 45 01.53 &   -77 14 46.6 &                G8III &       1 & $\mu$,$\pi$ &      2 &     8.41 &     0.54 &     0.16 &     5.74  \\
      F1 &   10 49 01.37 &   -76 54 27.9 &                K4III &       3 &    sp & 3,4 &     8.79 &     0.71 &     0.30 &     6.35  \\
 HD 94216$\tablenotemark{d}$ &   10 49 31.65 &   -77 06 33.6 &               F8IV/V &       1 &      $\mu$ &      5 &     8.81 &     0.22 &     0.09 &     7.86  \\
      F2 &   10 50 36.85 &   -77 58 05.1 &                  B8V &       3 &    sp & 3,4 &     9.33 &     0.09 &     0.08 &     8.50  \\
 HD 94414 &   10 50 55.92 &   -77 07 28.6 &                  B4V &       1 & sp,$\pi$ &  4,2 &     8.51 &     0.13 &     0.14 &     6.41  \\
      T1$\tablenotemark{d}$ &   10 52 01.34 &   -77 09 50.3 &                G3-G9 &        6 &    sp &      6 &    15.47 &     0.60 &     0.22 &    13.12  \\
IRAS 10529-7638$\tablenotemark{d}$ &   10 54 11.29 &   -76 54 36.6 &                M6III &        6 &    sp &      6 &     8.33 &     1.11 &     0.29 &     4.00  \\
      T2 &   10 54 13.30 &   -77 55 09.8 &          RR Lyrae,F-G &   7,6 & var,sp & 7,6 &    13.15 &     0.56 &     0.21 &    11.48  \\
\enddata
\tablenotetext{a}{2MASS Point Source Catalog.}
\tablenotetext{b}{Status as a background field star is indicated by 
$A_V\gtrsim1$ and either 
a spectral classification that places it below the main sequence
for the distance of Chamaeleon~I or that shows it to be a dwarf or giant 
(``sp"), weak Li absorption (``Li"), strong Na~I and K~I absorption (``NaK"),
type of variability (``var"), or a proper motion (``$\mu$"),
parallax (``$\pi$"), or radial velocity (``rv") that differs from that of
the known members of Chamaeleon~I.}
\tablenotetext{c}{Second DENIS Release.}
\tablenotetext{d}{Additional comments on this source are 
in \S~\ref{sec:append}.}
\tablerefs{
(1) \citet{hc75};
(2) \citet{per97};
(3) \citet{vr84};
(4) \citet{whi87};
(5) \citet{hog98};
(6) this work;
(7) \citet{wb96};
(8) \citet{hlf94};
(9) \citet{gm03};
(10) \citet{law96};
(11) \citet{fk89};
(12) \citet{wal92};
(13) \citet{saf03};
(14) \citet{gp02};
(15) \citet{kho85};
(16) \citet{whi91};
(17) \citet{rb88}.}
\end{deluxetable}

\begin{deluxetable}{lllllllllll}
\setlength{\tabcolsep}{0.04in}
\tabletypesize{\scriptsize}
\tablewidth{0pt}
\tablecaption{Sources with Uncertain Membership Status \label{tab:unc}}
\tablehead{
\colhead{ID} &
\colhead{$\alpha$(J2000)} &
\colhead{$\delta$(J2000)} &
\colhead{Ref} &
\colhead{Spectral Type} &
\colhead{Ref} &
\colhead{Status} &
\colhead{$i$\tablenotemark{a}} & 
\colhead{$J-H$\tablenotemark{b}} & \colhead{$H-K_s$\tablenotemark{b}} 
& \colhead{$K_s$\tablenotemark{b}} \\
\colhead{} &
\colhead{} &
\colhead{} &
\colhead{} &
\colhead{} &
\colhead{} &
\colhead{} &
\colhead{} &
\colhead{} &
\colhead{} &
\colhead{}
}
\startdata
     B23 &    11 01 46.6 &     -77 43 51 &  1 &                   \nodata &         \nodata &       ? &       \nodata &       \nodata &       \nodata &       \nodata  \\
  Cam2-8 &   11 02 46.76 &   -77 38 08.0 &   2 &                   M1 &       3 &       ? &       \nodata &     2.27 &     1.04 &    10.50  \\
 Cam2-12 &   11 04 10.81 &   -77 50 13.1 &   2 &                   M0 &       3 &      background? &    18.48 &     2.36 &     1.00 &    10.59  \\
     B29 &    11 05 01.9 &     -77 41 31 &  1 &                   \nodata &         \nodata &       ? &       \nodata &       \nodata &       \nodata &       \nodata  \\
 Cam2-14 &   11 05 18.88 &   -77 57 38.9 &   2 &                 M0.5 &       3 &      background? &    18.00 &     2.21 &     1.00 &    10.53  \\
 Cam2-18 &   11 05 54.54 &   -77 35 12.3 &   2 &             M0.5,$<$M0 &   3,4 &      background? &    17.98 &     2.32 &     1.06 &     9.96  \\
     B31 &    11 06 26.4 &     -77 34 20 &  1 &                   \nodata &         \nodata &       ? &       \nodata &       \nodata &       \nodata &       \nodata  \\
 Cam2-33 &   11 08 11.62 &   -77 18 53.4 &   2 &                 M0.5 &       3 &      background? &       \nodata &     2.71 &     1.23 &    10.57  \\
\enddata
\tablenotetext{a}{Second DENIS Release.}
\tablenotetext{b}{2MASS Point Source Catalog.}
\tablenotetext{c}{Additional comments on this source are 
in \S~\ref{sec:append}.}
\tablerefs{
(1) \citet{car02};
(2) 2MASS Point Source Catalog;
(3) \citet{gm03};
(4) this work;
(5) \citet{vr84};
(6) \citet{fk89}.}
\end{deluxetable}

\clearpage

\clearpage

\clearpage

\clearpage

\clearpage

\clearpage

\clearpage

\clearpage

\clearpage

\clearpage

\clearpage

\clearpage

\clearpage

\clearpage

\clearpage

\clearpage

\clearpage

\clearpage

\clearpage

\clearpage

\clearpage

\clearpage

\clearpage

\end{document}